\documentclass[letterpaper]{article} % DO NOT CHANGE THIS
\usepackage{aaai2026}  % DO NOT CHANGE THIS
\usepackage{times}  % DO NOT CHANGE THIS
\usepackage{helvet}  % DO NOT CHANGE THIS
\usepackage{courier}  % DO NOT CHANGE THIS
\usepackage[hyphens]{url}  % DO NOT CHANGE THIS
\usepackage{graphicx} % DO NOT CHANGE THIS
\usepackage{natbib}  % DO NOT CHANGE THIS AND DO NOT ADD ANY OPTIONS TO IT
\usepackage{caption} % DO NOT CHANGE THIS AND DO NOT ADD ANY OPTIONS TO IT
\usepackage{algorithm}
\usepackage{algorithmic}

\usepackage{newfloat}
\usepackage{listings}
\DeclareCaptionStyle{ruled}{labelfont=normalfont,labelsep=colon,strut=off} % DO NOT CHANGE THIS
\floatstyle{ruled}
\newfloat{listing}{tb}{lst}{}
\floatname{listing}{Listing}
\usepackage{xcolor}
\usepackage{booktabs}
\usepackage{placeins}
\usepackage{makecell}
\usepackage[most]{tcolorbox}
\newcommand{\answerYes}[1]{\textcolor{blue}{#1}} 
\newcommand{\answerNo}[1]{\textcolor{teal}{#1}} 
\newcommand{\answerNA}[1]{\textcolor{gray}{#1}}

\title{From Pixels to Personas: Tracking the Evolution of Anime Characters}
\author {
    Rongze Liu\textsuperscript{\rm 1},
    Jiaxin Pei\textsuperscript{\rm 2,3},
    Jian Zhu\textsuperscript{\rm 1}
}
\affiliations {
    \textsuperscript{\rm 1}University of British Columbia\\
    \textsuperscript{\rm 2}Stanford University\\
    \textsuperscript{\rm 3}University of Texas at Austin\\
    celt313@student.ubc.ca, pedropei@stanford.edu, jian.zhu@ubc.ca
}

\begin{document}

\maketitle

\begin{abstract}
Anime, originated from Japan, is one of the most influential cultural products in modern society and is especially popular among younger generations. The popularity of anime reflects important cultural evolutions in our society. Despite existing research on anime as a cultural phenomenon, we still have a limited understanding of how anime really evolves over the years. In this study, using a large-scale multimodal dataset of anime characters from an anime review site, we applied computational methods that integrate textual, visual, and production features of anime characters with online popularity traces. By combining LLM-extracted personality features with avatar features, we identify recurring personality archetypes and visual tropes with their temporal evolution over the past decades. We found that the target audience of anime has undergone a systematic shift from children to a maturing audience of teenagers and young adults over time. Character design has been undergoing \textit{moe}-ification, with softer or sexualized female traits becoming increasingly prominent since the 2000s. Some personality archetypes are often visually predictable, yet audiences also tend to prefer less conventionalized characters. Finally, we reveal that visual signals play a more dominant role than personality traits in shaping audience preferences, with features such as \textit{moe}-style faces and mechanical designs contributing greatly to popularity. These findings offer insights into the broader dynamics of anime's cultural and creative practices.
\end{abstract}
\begin{links}
    \link{Code}{https://github.com/Arc-Celt/pixels2personas}
\end{links}

\section{Introduction} 

Anime, together with Manga and video games, constitute the most important cultural products of contemporary Japan \cite{oohagan2007manga}. Artistically, anime and Manga represent a distinct visual style and aesthetics in contemporary Japanese culture and national image \cite{norris200913}. Anime has a profound impact on the popular cultures in East Asia, and the rest of the world, giving rise to spin-off merchandise like figurines  \cite{condry2009anime}, anime tourism \cite{okamoto2015otaku}, Internet memes \cite{saito2017moe}, etc., and new digital forms like VTubers \cite{10.1145/3706598.3713877}. In 2025, Netflix reports that fifty percent of global viewers on its own platform now watch anime \cite{brzeski2025netflix}. 
Despite its global impact, anime has received surprisingly little attention in digital humanities and computational social sciences,  with very few studies focusing on the computational analysis of anime and manga \cite{douglass2011understanding,manovich2012compare}, as compared to other media like movies, music, and games.

Unique and appealing characters are one of the most long-lasting attractions in anime. In prior research, characters in Japanese manga and anime are often referred to as ``\textit{kyara}" (Japanese transliteration of characters), to distinguish them from conventional characters \cite{hosei2023prince,go2011tezuka}. \textit{Kyara} is a character that is identifiable even beyond the original story context. Such characters can persist beyond anime, manga, and video games, in merchandise like figurines, key chains, and CDs, and in fan-made works like fan fiction and cosplay. However, as hundreds of characters are churned out by the anime industry every year, not all of them will persist beyond the airing, except for a few preferred by the audience. 

\subsubsection{Research Questions}
We conceptualize anime characters as cultural products with multimodal design elements encompassing personality traits, distinct visual features, and elaborate voice acting. Anime has been easily recognizable as being `anime-esque' for its distinguishing art style \cite{suan2017anime}. Even within the surface uniformity of anime style, there still exists a large space for unique character design. While we know that character design matters \cite{iijima2016proportion}, we lack large-scale, data-driven evidence of how personality and visual style capture audience attention. By utilizing the online crowd-sourced preference data from \texttt{MyAnimeList}, an anime review site, our study progresses from macro to micro: first examining historical production trends to establish the evolution context, then identifying dominant character design patterns across periods, and finally exploring how these design elements are systematically associated with audience preference. Specifically, we ask the following research questions:

\begin{itemize}
    \item \textbf{RQ1:} What is the historical trend of anime production?
    \item \textbf{RQ2:} What are the most common character archetypes in specific periods in the past decades? 
    \item \textbf{RQ3:} What are the common visual tropes in character creation? How do the artistic styles of anime characters change over time?
    \item \textbf{RQ4:} Are there any systematic associations between personality traits and visual features?
    \item \textbf{RQ5:} How do personality and visual styles predict (mostly Western) audience preference, as measured by crowd-sourced favorite counts?
\end{itemize}

\subsubsection{Contributions}
To answer these research questions, we curated a comprehensive anime database, covering over 20k animes in the past decades and over 150k unique characters. Large-scale computational methods were applied to analyze the temporal trends of anime characters. Our study makes the following contributions. 

\begin{itemize}
    \item \textbf{We uncovered core personality archetypes and recurring visual tropes, tracing their historical evolution and revealing their associations.} Our results reveal that character personality archetypes have shifted toward greater psychological complexity, indicating a shift toward a maturing audience of teenagers and young adults. Anime character design has been significantly influenced by \textit{moe}-ification, the design trend targeting \textit{moe} (affectionate responses to fictional characters) \cite{galbraith2014moe}, with soft or sexualized female visual traits becoming increasingly prevalent during recent decades. Moreover, personality archetypes are often visually predictable, especially for certain flat archetypes, suggesting that personality and visual features operate symbiotically for coherent character design.
    \item \textbf{We demonstrated how personality and artistic designs jointly shape audience preferences within Japanese anime's established design conventions.} While both artistic styles and personality traits contribute to audience appeal, visual features exhibit greater predictive power. Characters that feature feminine traits like \textit{moe}-style faces, long stylish hair, and slender proportional bodies, as well as elaborate mechanical designs like armor or \textit{mecha} elements, tend to achieve higher popularity. This illustrates how anime's cultural production strikes a balance between creative innovations and cultural conventions through mixing and matching of recognizable anime-esque elements \cite{suan2017anime}.
\end{itemize}

\section{Related Works}

\subsection{Anime}
Anime is often created not only for its own sake but as part of the interconnected cultural products. Media mix is a commercial strategy adopted by the Japanese companies to develop transmedia franchises that span anime, manga, video games, spin-off toys, and other franchised toys \cite{steinberg2012anime,kopylova2023drawing,napier2011manga,kinoshita2024gundam,steinberg2023introducing}. For example, the \textit{Mobile Suit Gundam} franchise extends its TV anime series with novelizations, model kits, and life-sized statues, which help sustain and expand its cultural impact \cite{kinoshita2024gundam}. Such a strategy also exerts an influence on the visual style of anime character design \cite{kopylova2023drawing,watzky2023dynamism}.
Many analyses or overviews of Japanese anime tend to focus exclusively on a few influential hits like those from the Ghibli Studio \cite{lunning2011under}. Yet the scope has rarely been expanded to include wider ranges of anime created across decades. 

\subsection{Anime Characters}
The artistic style of anime characters has been investigated in the humanities and social sciences, covering the creation pipeline \cite{condry2009anime}, racial perception \cite{lu2009race,fennell2013consuming}, popularity \cite{tang2024correlation}, voice dubbing \cite{howell2007character}, parasocial attachment \cite{ramasubramanian2012japanese} and cosplay \cite{winge2006costuming}. Some character design has also been criticized for pandering to the \textit{Otaku} community, a subculture of devoted anime and manga fans \cite{azuma2009otaku} that are mostly heterosexual males, perpetuating female stereotypes \cite{ting2019gender,lunning2011under}. Yet, some anime, like \textit{Sailor Moon} series, once regarded as objectifying, was also argued to subvert patriarchal and gender stereotypes \cite{yatron202230}. 

Anime characters have been argued to be \textit{Mukokuseki} (meaning `nationless' or `stateless'), as the physical appearance of many characters does not look Japanese but more like abstract human bodies, depending on the creators and genre tropes \cite{ruh2014conceptualizing}. Many classic anime characters will also become the foundation for the creation of new characters for the \textit{Otaku} community to consume, referred to as `database consumption' \cite{azuma2009otaku}, or `database fantasyscape' \cite{ruh2014conceptualizing}. While the interconnectivity of anime character design and tropes has long been recognized by scholars, it is mostly approached from a qualitative perspective with case studies and impressionistic analyses. Such analyses can be complemented by large-scale computational analysis, which can serve to augment the discovery of shared cultural resources in the design of anime characters. 

Computational modelling of anime characters in the Computer Vision Communities tend to focus on generating anime avatars from text or image prompts through training on large-scale datasets \cite[e.g.,][]{jin2017towards,li2021anigan,huang2023anifacedrawing,cao2023animediffusion}. Neural networks have also been trained to recognize the distinct artistic style of anime characters across different anime works \cite{li2022challenging}. However, these computational studies only center on predictions through neural networks that are usually uninterpretable. There still lacks a comprehensive study to understand the design and the cultural affordance of anime character design.

\section{Method}
\subsection{Data Collection}
The anime data was gathered from the anime review site \texttt{MyAnimeList}\footnote{\url{https://myanimelist.net/}}, one of the most comprehensive websites for anime. Each month, the site is visited by over 12 million unique users worldwide (75\% male), encompassing 200 countries across the globe (25\% from the US)\footnote{\url{https://myanimelist.net/advertising}}.

Data were scraped and parsed with customized scripts.
In addition to the anime main pages, we further retrieved the pages for each individual character, if available. These individual character pages usually contain textual descriptions of anime characters with varying details, contributed by community members or other public sources like Wikipedia. We also retrieved the high-resolution character avatars if available. We parsed the crawled webpages to extract meta information (years and user ratings), and text descriptions of anime and characters. Voice is not analyzed in this work due to the lack of audio data. 
All anime shows aired after July 1, 2025 were excluded. We adhered to ethical practices of data collection. The anime database only contains public data without sensitive information. We set an interval of 10 seconds between consecutive requests to avoid burdening the website server. The whole crawling process took around two weeks to complete.

\subsection{Analyzing Character Personalities}
\subsubsection{Character Labeling}
Anime characters on \texttt{MyAnimeList} are usually accompanied by unstructured free text descriptions. LLMs have also been shown to exhibit strong capabilities to understand literary characters \cite{jang-jung-2024-evaluating,amalvy-etal-2025-role,papoudakis-etal-2024-bookworm,abrams2025literary}. So we prompted an LLM to summarize personality traits in a few adjectives, given the crowd-sourced character descriptions on \texttt{MyAnimeList}. Sample prompts and outputs are available in the Appendix.

We utilized Qwen3 \cite{yang2025qwen3}, a series of state-of-the-art open-sourced LLMs, to systematically annotate the character descriptions into structured representations. The \texttt{Qwen3-32B-FP8} was deployed through \texttt{vLLM} \cite{kwon2023efficient} in \texttt{FP8} precision, and the model performed inference in non-thinking mode. 

\subsubsection{Model Validation}

To validate our model outputs, we randomly sampled 50 character instances. For each instance, we created a candidate pool of 21 shuffled keywords: the model-generated personality keywords for that character, distractor keywords randomly sampled from other characters' keyword sets, and a “None of the above” option. Using the Potato annotation tool \cite{pei2022potato}, two of the authors annotated all 50 samples independently, with no prior knowledge of the samples. For each character, annotators were asked to read the biography and select all keywords from the candidate pool that accurately described the character's personality. We then measured the agreement between human selections and model-generated keywords using metrics of precision, recall, F1-score, and Jaccard similarity.

\begin{table}[ht]
    \centering
    \begin{tabular}{lcccc}
    \toprule
    \textbf{Annotator} & \textbf{Precision} & \textbf{Recall} & \textbf{F1} & \textbf{Jaccard} \\
    \midrule
    \makecell[c]{1} & 0.907 & 0.671 & 0.747 & 0.639 \\
    \midrule
    \makecell[c]{2} & 0.767 & 0.656 & 0.688 & 0.566 \\
    \bottomrule
    \end{tabular}
    \caption{Mean agreement metrics between LLM-extracted personality keywords and human annotations for 50 random samples: Precision (fraction of model keywords validated by humans), Recall (fraction of human-selected keywords generated by the model), F1-score, and Jaccard similarity (intersection over union of selected keyword sets). They demonstrate substantial alignment between LLM extractions and human judgments. }
    \label{tab:human_evaluation_results}
\end{table}

As shown in Table~\ref{tab:human_evaluation_results}, both annotators achieved high agreement with the model-generated keywords, as indicated by the high average agreement scores across all metrics. This consistency suggests that the model outputs are generally reliable and align well with human judgments.

\subsubsection{Clustering Character Personalities}
We extracted the 1024-dimensional embeddings for the personality traits of each character using \texttt{Qwen3-Embedding-0.6B} \cite{zhang2025qwen3}. Each embedding vector was L2-normalized to unit length. Then we applied Uniform Manifold Approximation and Projection (UMAP) \cite{mcinnes2018umap}, a nonlinear dimensionality reduction technique, to reduce the 1024-dimensional vectors to two dimensions.

We constructed a distance-thresholded graph where nodes represent characters and edges connect characters within a specified distance threshold (5th percentile). Edge weights were computed as $w_{ij} = 1/(1 + d_{ij})$, where $d_{ij}$ is the Euclidean distance between characters in the UMAP space. To identify groups of characters with similar personality profiles, we performed community detection on the 2D UMAP embeddings using the Leiden algorithm \cite{Traag_2019}, yielding a partition with high modularity. The top 10 nodes with the highest degree centrality were selected as representatives of the personality archetype associated with their community.

\subsection{Analyzing Avatars}
\subsubsection{Avatar Processing} We extracted the latent representations of character avatars with DINOv2 \cite{oquab2024dinov2}, a vision transformer (ViT) encoder model \cite{dosovitskiy2020image} pretrained on large-scale images with self-supervision. DINOv2 is capable of extracting fine-grained visual features from raw images as represented by embedding vectors for many downstream tasks like image classification and clustering \cite{oquab2024dinov2}. Specifically, we used \texttt{dinov2-base}\footnote{\url{https://huggingface.co/facebook/dinov2-base}}, an 82M parameter variant for all feature extractions. All original avatar images were rescaled to $224\times224$ to fit DINOv2's native resolution, and then all rescaled images were transformed into 768-dimensional DINOv2 embeddings. 

\subsubsection{Avatar Tagging} Genders of characters are not explicitly labeled in the character page. So we utilized the \texttt{Camie Tagger}\footnote{\url{https://huggingface.co/Camais03/camie-tagger}}, a state-of-the-art anime image tagger that has been trained on over 7 million anime images with over 70k tags.
We extracted the gender tags from the predicted outputs. Each avatar is labeled as male, female, or other (for non-binary or non-humanoid characters).

\subsubsection{Finetuning DINOv2}

To investigate the temporal changes in character art style design, we finetuned DINOv2 to classify the years of the characters. All images were labeled according to their anime's release year into 10 temporal periods, then partitioned with stratified sampling as 40\% for training, 10\% for validation, and 50\% for testing.

We appended a 3-layer MLP classifier to the DINOv2 backbone with ReLU activation and dropout regularization. As DINOv2 was pretrained on a vastly different visual domain than anime avatars\cite{oquab2024dinov2}, we implemented progressive unfreezing: train the head only first, then unfreeze the last 2 blocks at epoch 8, and expand to the last 4 blocks at epoch 12.

\begin{table}[t]
    \centering
    \begin{tabular}{ll}
    \toprule
    Hyperparameter & Value \\\midrule
    Training Epochs & 200 \\
    Min Epochs & 10 \\
    Learning Rate & 3e-5 \\
    Batch Size & 20 \\
    Weight Decay & 0.05 \\
    Early Stopping Patience & 10 \\
    \bottomrule
    \end{tabular}
    \caption{Fine-tuning hyperparameters for \texttt{dinov2-base}.}
    \label{table:dinov2_finetuning_hyperparams}
\end{table}

\section{RQ1: Historical Trends in Anime}
\begin{figure}[t]
    \centering
    \includegraphics[width=0.97\columnwidth]{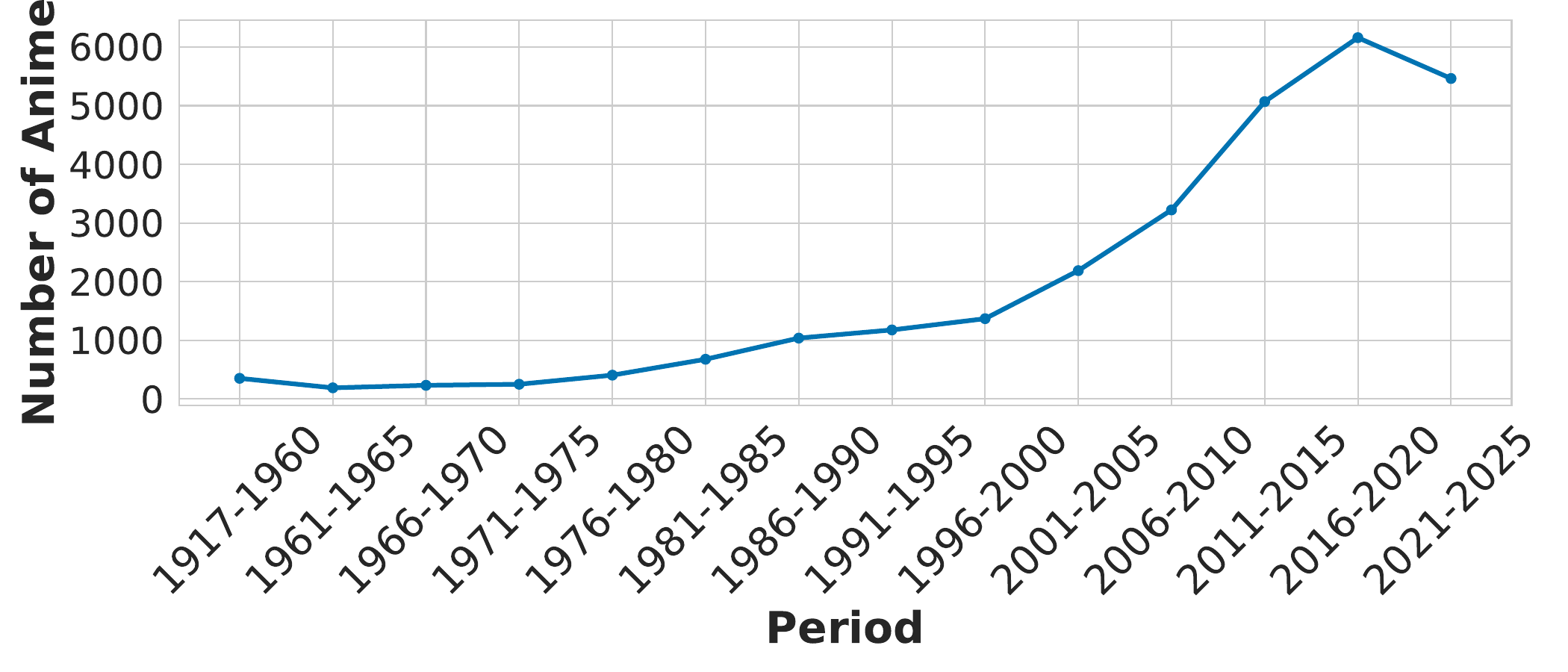}
    \includegraphics[width=0.97\columnwidth]{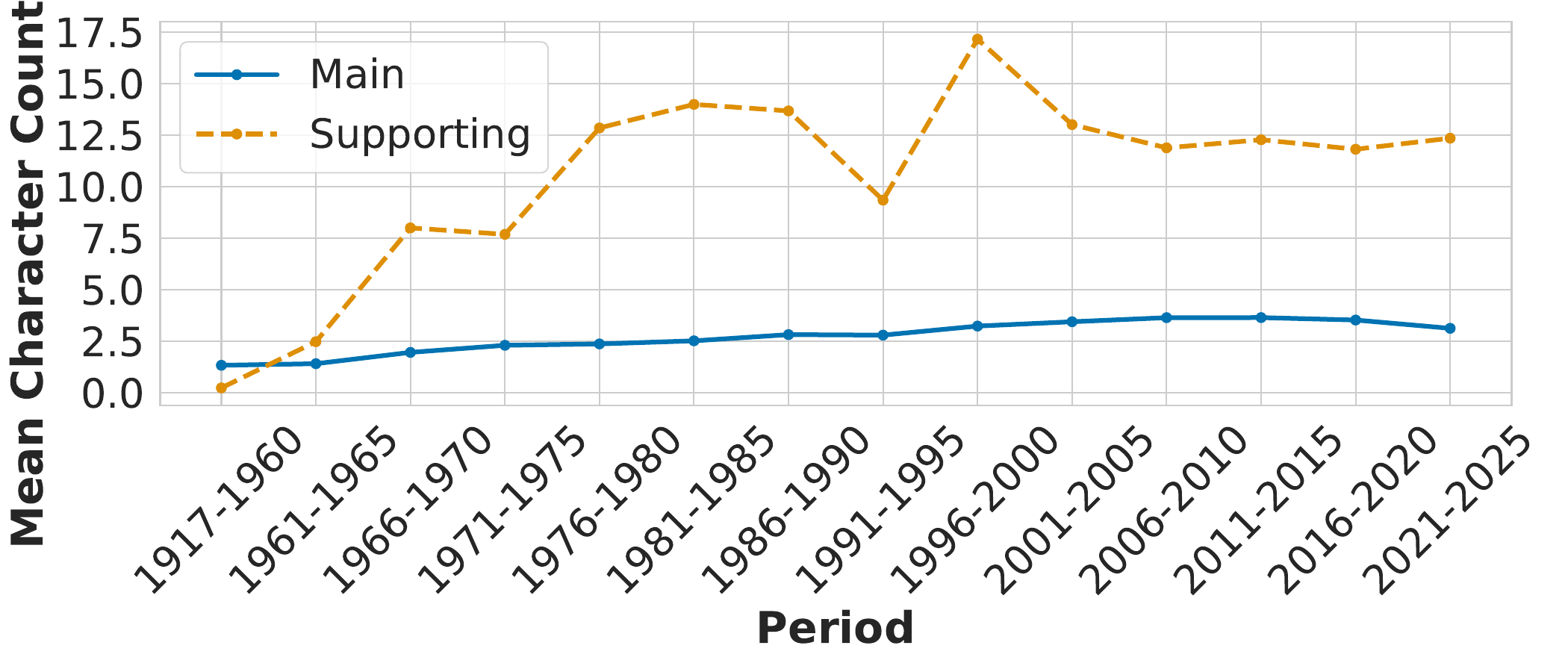}
    \caption{Number of anime per period (top) and average number of characters per anime over time (bottom). Anime production has grown rapidly, with supporting characters steadily increasing, indicating richer story structures and more complex narratives.}
    \label{fig:anime-chars-desc}
\end{figure}

\textbf{Generally speaking, anime production has expanded significantly historically, while the medium itself has grown more mature, more complex, and has benefited from a wider range of sources}. Table~\ref{table:general_stats} summarizes the statistics of our dataset. In general, we have collected over 27,000 animes with over 130k characters, spanning 108 years. We can also observe several general trends in anime production. In Figure~\ref{fig:anime-chars-desc}, the number of anime produced per year has increased steadily since the 1950s, but the increase picked up momentum around the 2000s and peaked after 2015. Together with this increase, the structures of anime have also become richer and more complex in recent decades. While the number of main characters barely changes, the number of supporting characters has undergone a steady increase. This trend may stem from the modernization of the anime industry, which enables larger supporting casts, as well as market shifts toward more complex storytelling.

\begin{table}[t]
\centering
\begin{tabular}{ll}
\toprule
Fields & Values \\\midrule
Number of Anime            &              27,783              \\ 
Number of Main Characters       &             26,722               \\ 
 - with text descriptions       &             18,002               \\ 
Number of Supporting Characters       &               111,777             \\ 
 - with text descriptions       &              54,274             \\ 
Number of Avatars             &        126,347       \\
%Number of Genres                 &              21              \\ 
%Number of Studios                &              1,240              \\ 
%Average Length of Anime (episodes) &              13.23              \\ 
Release Year Range               &              1917 - 2025              \\ \bottomrule
\end{tabular}
\caption{General Dataset Statistics}
\label{table:general_stats}
\end{table}

\begin{figure}[t]
    \centering
    \includegraphics[width=0.97\columnwidth]{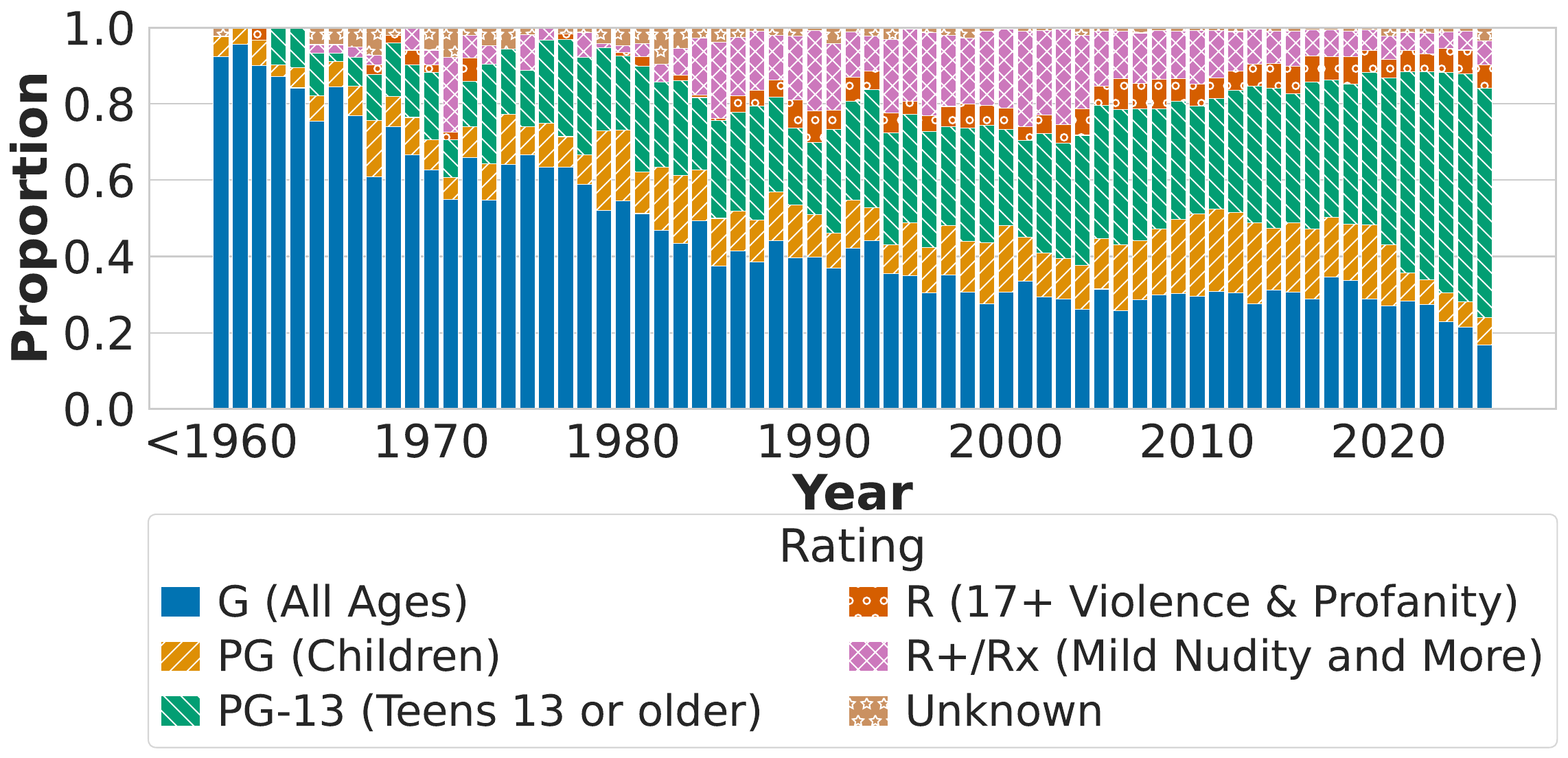}
    \caption{Proportion of PG ratings by year. The decline in G-rated anime alongside the substantial rise in PG-13 content demonstrates a clear shift toward more mature audiences.}
    \label{fig:pg-rating-year}
\end{figure}

Figure~\ref{fig:pg-rating-year} suggests that the target audience for anime has shifted toward a more adult demographic over the past few decades. The most notable trends include a rapid increase in the number of PG-13–rated anime and a corresponding decline in G-rated (All Ages) and PG-rated (Children) titles. Additionally, anime rated R and R+ experienced a short period of growth during the 2000s but began to decline after around 2010. These changes might reflect a maturing audience, and are likely also fueled by the shifting production model by streaming platforms like Netflix, which allows for adult-like content oriented at a global audience \cite{noh2024global}.

\section{RQ2: Character Personality Archetypes}
We investigate recurring patterns in character personality designs by detecting common archetypes, analyzing their role and gender distributions, and tracing temporal trends.

\subsection{Identifying Character Archetypes}

\begin{table*}[t]
\centering
\begin{tabular}{llll}
\toprule
Archetype & Main / Supporting & Traits & Characters\\
\midrule
Energetic Extroverts   & 44.4\% / 55.6\% & cheerful, energetic, outgoing & Mini Yaemori, Gokuu Son \\
Steadfast Leaders      & 32.2\% / 67.8\% & calm, responsible, leader-like & Roy Mustang, Saber \\
Cold-hearted Antagonists        & 22.5\% / 77.5\% & cold-hearted, arrogant, cunning & Dio Brando, Frederica Bernkastel \\
Passionate Strivers           & 36.3\% / 63.7\% & passionate, dedicated, determined & Kyoujurou Rengoku, Aaeru \\
Kind Caregivers      & 36.1\% / 63.9\% & kind, compassionate, considerate  & Rem, Haruno Yukinoshita \\
Justice Keepers      & 41.5\% / 58.5\% & sense of justice, loyal & Setsuna F. Seiei, Convoy \\
Reserved Introverts       & 61.7\% / 38.3\% & introverted, timid, reserved & Mio Akiyama, Sakura Adachi \\
Arrogant Tsunderes     & 50.3\% / 49.7\% & tsundere, arrogant, sarcastic & Mikoto Misaka, Taiga Aisaka \\
\bottomrule
\end{tabular}
\caption{Character Personality Archetypes.}
\label{table:char_personality_proto}
\end{table*}

Based on the Leiden community detection results on the UMAP-reduced personality embeddings, \textbf{we are able to cluster all characters into 8 major personality archetypes that recur in anime}.  Each archetype corresponds to a cluster of characters sharing similar personality profiles, characterized by the top 10 most central characters and their dominant personality keywords. Table~\ref{table:char_personality_proto} summarizes these archetypes and their role distributions.

\textbf{Anime creators prefer different archetypes for main and supporting characters, with main characters tending to be more complex and multidimensional}. Reserved Introverts, Arrogant Tsunderes, Energetic Extroverts, and Justice Keepers are more likely to be main characters than other archetypes. The Energetic Extroverts and Arrogant Tsunderes are common archetypes in \textit{Shonen} (teenagers) manga and anime; yet, characters with intense emotional conflicts and internal activities have also become increasingly prominent. These archetypes allow for character growth throughout the anime and provide stronger self-identification with the audience. Comparatively, supporting characters tend to be Cold-hearted Antagonists, Steadfast Leaders, Kind Caregivers, and Passionate Strivers. These archetypes tend to be flat characters by design.

\subsection{Archetype Patterns by Gender}

\begin{figure}[t]
    \centering
        \includegraphics[width=0.97\columnwidth]{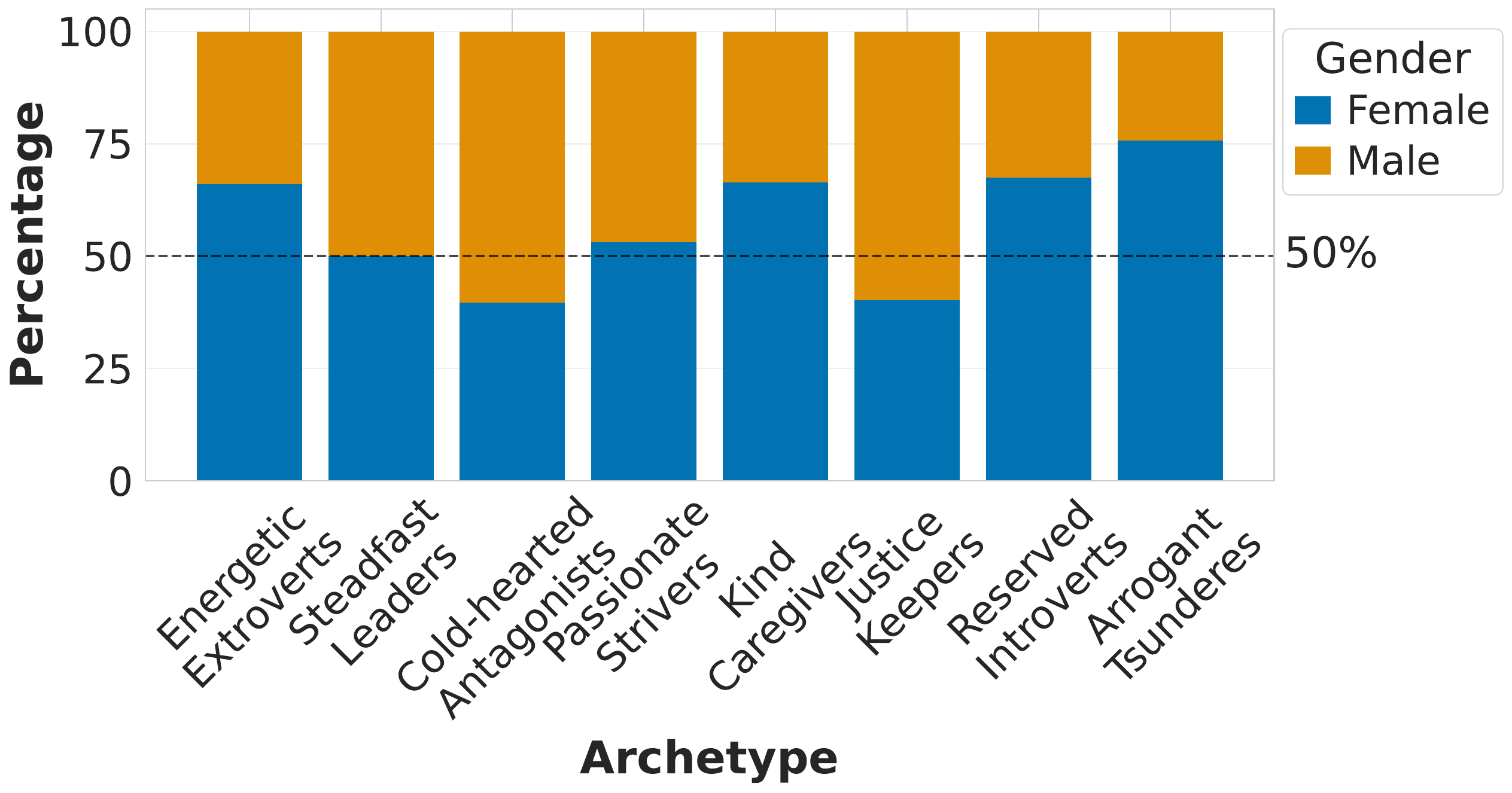}
        \caption{Gender share within each archetype. Female characters are modestly overrepresented in Arrogant Tsunderes, Reserved Introverts, Energetic Extroverts, and Kind Caregivers, whereas Cold-hearted Antagonists and Justice Keepers are more frequently associated with males.} % Archetype share across rating categories. More mature anime features more complex or darker archetypes.}
        \label{fig:community-vs-gender}
\end{figure}

Figure~\ref{fig:community-vs-gender} illustrates how personality archetypes are distributed across gender. Arrogant Tsunderes, Reserved Introverts, Energetic Extroverts, and Kind Caregivers are predominantly female, while archetypes such as Cold-hearted Antagonists and Justice Keepers have a higher proportion of male characters. \textbf{Anime storytelling is shaped by traditional gender norms, where males are consistently depicted as tenacious, dominant, and competent, while females are more often portrayed as dependent, emotional, and supportive} \cite{Bresnahan2006}.

\subsection{Temporal Changes in Character Personality Design}

\begin{figure*}[htbp]
    \centering
        \includegraphics[width=0.85\textwidth]{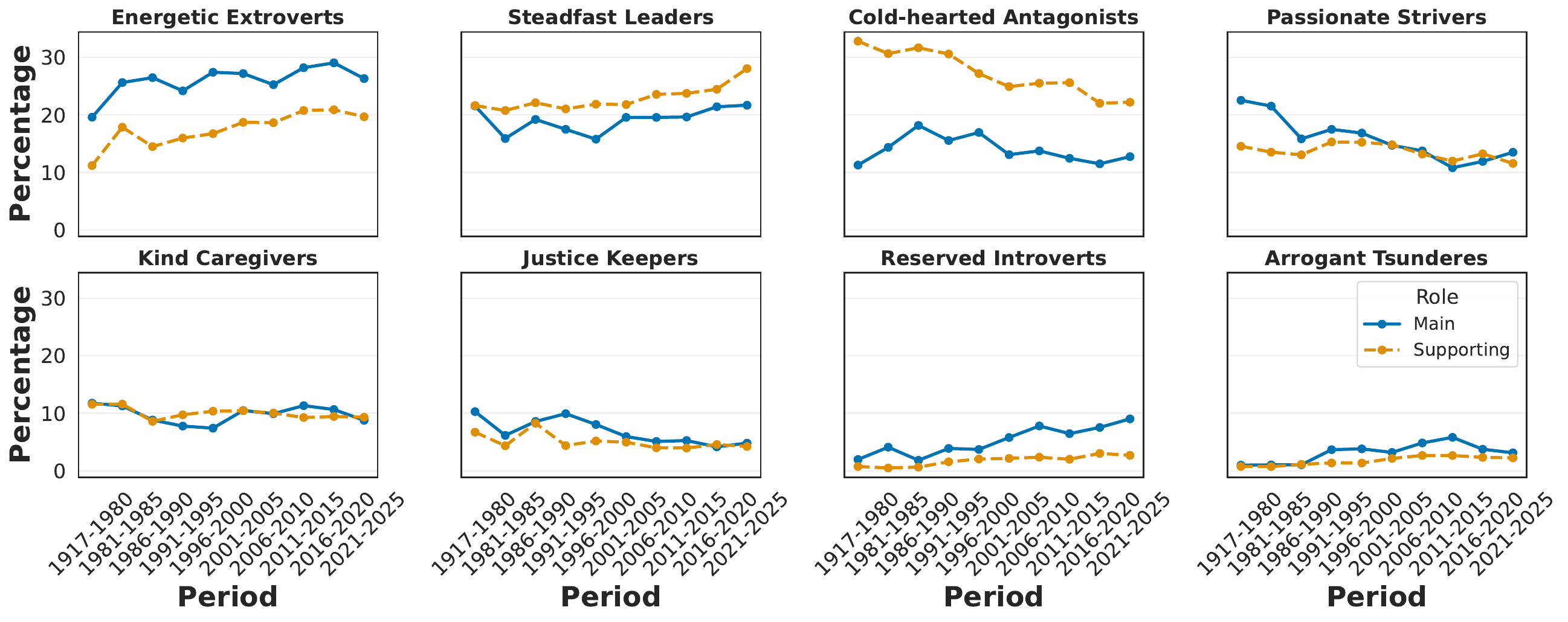}
        \caption{Temporal trends of personality archetypes among anime characters by role. Main characters have consistently favored active and outgoing types, though psychologically complex archetypes like Reserved Introverts and Arrogant Tsunderes have risen modestly. Supporting characters have become more diverse, with fewer antiheroes and more individualized personalities.}
        \label{fig:temporal-change-linechart}
\end{figure*}

\textbf{The distribution of personality archetype design has remained generally stable over time, with minor rise and fall of certain archetypes.} In Figure~\ref{fig:temporal-change-linechart}, among the main characters, Energetic Extroverts and Passionate Strivers have consistently been dominant, with slight fluctuations, reflecting the enduring demand for determined and outgoing protagonists who actively confront challenges and inspire others with their enthusiasm. Noticeably, however, the share of Reserved Introverts has gradually increased, especially in recent decades. This indicates a growing appreciation for psychologically complex protagonists, who may not possess the traits of classic heroes, but are flawed and constantly growing, making them more relatable to audiences \cite{tang2024correlation}. Arrogant Tsunderes, though a small group among main characters, are on the rise. Such characters are both resistant and appealing, with this contrast becoming increasingly prominent in character designs.

The distribution of archetypes among supporting characters also reveals several notable trends. Cold-hearted Antagonists have consistently occupied the largest share over time, but their prevalence has generally decreased over the decades. This suggests a pattern similar to that of protagonists: while anime still relies on antagonistic forces to drive conflict and plot, supporting characters are now depicted with greater diversity and complexity, surpassing traditional villain figures. Support-oriented roles such as Steadfast Leaders and Kind Caregivers have maintained a stable presence over time. They often serve as the steady backbone of a cast, playing a crucial role in aiding protagonists and fostering their growth. Meanwhile, the increase in both Energetic Extroverts and Reserved Introverts indicates a need for enriched interactions and contrasting perspectives, which help drive the protagonists' growth.

\section{RQ3: Character Avatar Archetypes}
Visual conventions in character avatars are analyzed through principal component analysis (PCA) on avatar embeddings and interpretation of the top components, alongside their temporal shifts.

\begin{figure}[htbp]
  \centering
  \includegraphics[width=0.97\columnwidth]{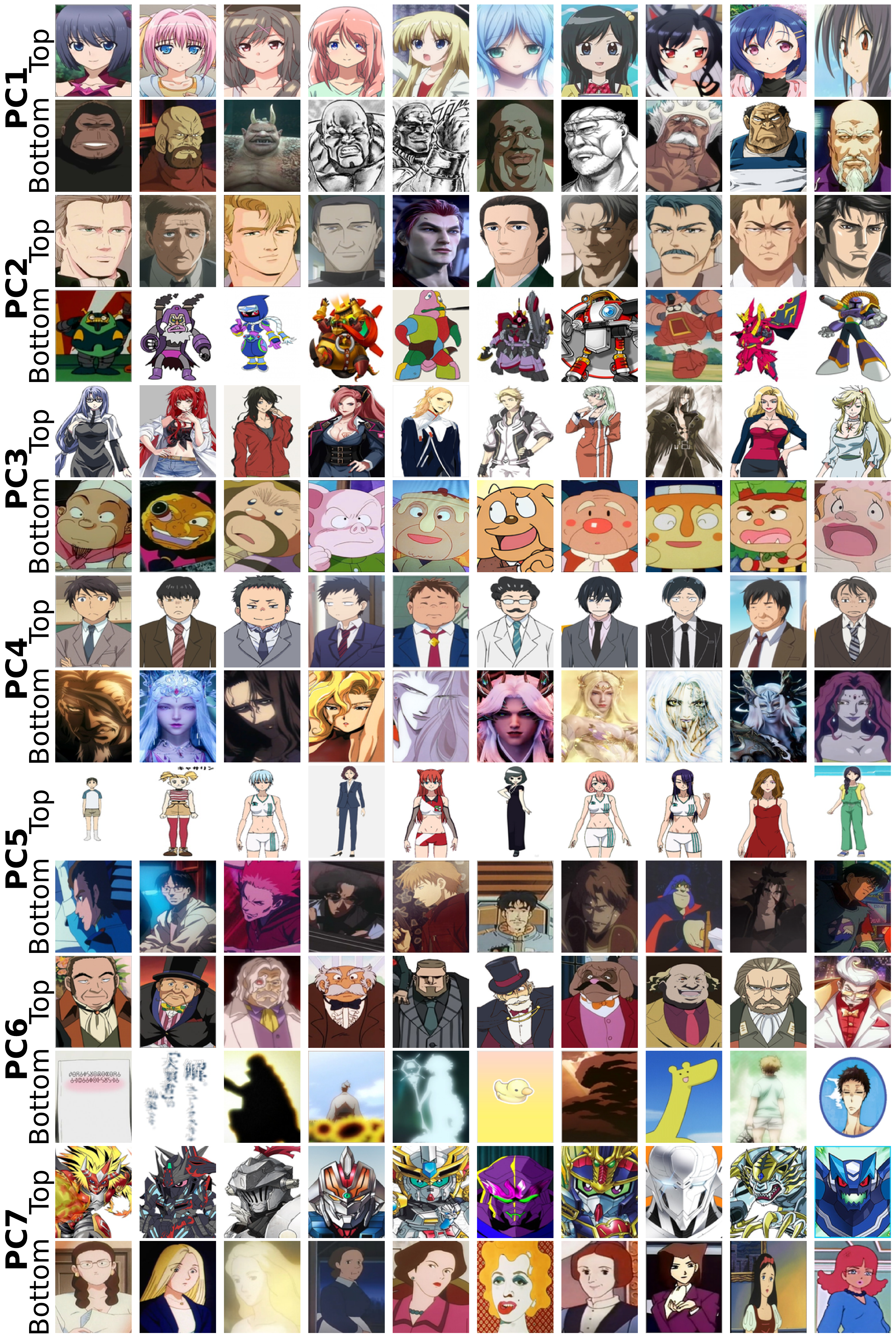}
  \caption{Examples of the top 10 (most positive) and bottom 10 (most negative) characters for each of the top 7 principal components (PCs) derived from avatar embeddings.}
  \label{fig:pc-image-example}
\end{figure}

\subsection{Dimensions of Character Design}

\textbf{Anime characters are often designed with recurring visual tropes}. We decomposed visual character design into interpretable orthogonal dimensions by applying PCA on the DINOv2 embeddings of character avatars. As reflected in the top and bottom character examples in Figure~\ref{fig:pc-image-example}, each principal component captures a distinct aspect of character design, which can be interpreted by examining the stylistic features of characters at both extremes. PC1 contrasts large, muscular males with fierce expressions and gentle, smiling females, representing a spectrum from aggressive, masculine features to approachable, feminine ones. PC2 separates realistic human males with angular facial features from flat, non-humanoid characteristics, spanning from realistic human forms to abstract or mechanical designs. PC3 distinguishes tall, slender figures with angular faces from round, cartoonish designs, capturing maturity versus childlike proportions. PC4 contrasts ordinary office workers with goddess-like fantasy figures, representing a spectrum from everyday to fantastical designs. PC5 differentiates bright, vibrant portrayals of young females in light clothing and simple settings from dark, shadowed depictions of serious male characters, reflecting a transition from luminous visuals to more intense and dark visual styles. PC6 contrasts detailed portraits of mature males with mustaches and formal attire with highly abstract avatars, including silhouettes, blurred figures, and animals, capturing a spectrum from recognizable characters to non-character visual elements. PC7 distinguishes armored, masked characters with sharp, angular designs from rounded, vintage-style female faces, representing a spectrum from mechanical, masked figures to classic, old-fashioned face designs.

\subsection{Temporal Changes in Art Styles}

\textbf{The art styles of anime characters have shifted systematically over time, with strong trends of \textit{moe}-ification}. Character avatars were split using stratified sampling (40\% training, 10\% validation, 50\% testing) and used to finetune DINOv2 for predicting character periods. The F1 score on the test set is 0.39, substantially outperforming the majority vote baseline (F1 = 0.03). The confusion matrix in Figure~\ref{fig:period-predict-confusion-mat} in the appendix demonstrates that temporal periods of anime character avatars are distinguishable by visual features. While each period has its own distinctive features, there is also a clear sense of continuity in the development of anime character art over time, as reflected by the fact that most errors occur between adjacent periods.

\begin{figure}[t]
    \centering
        \includegraphics[width=0.97\columnwidth]{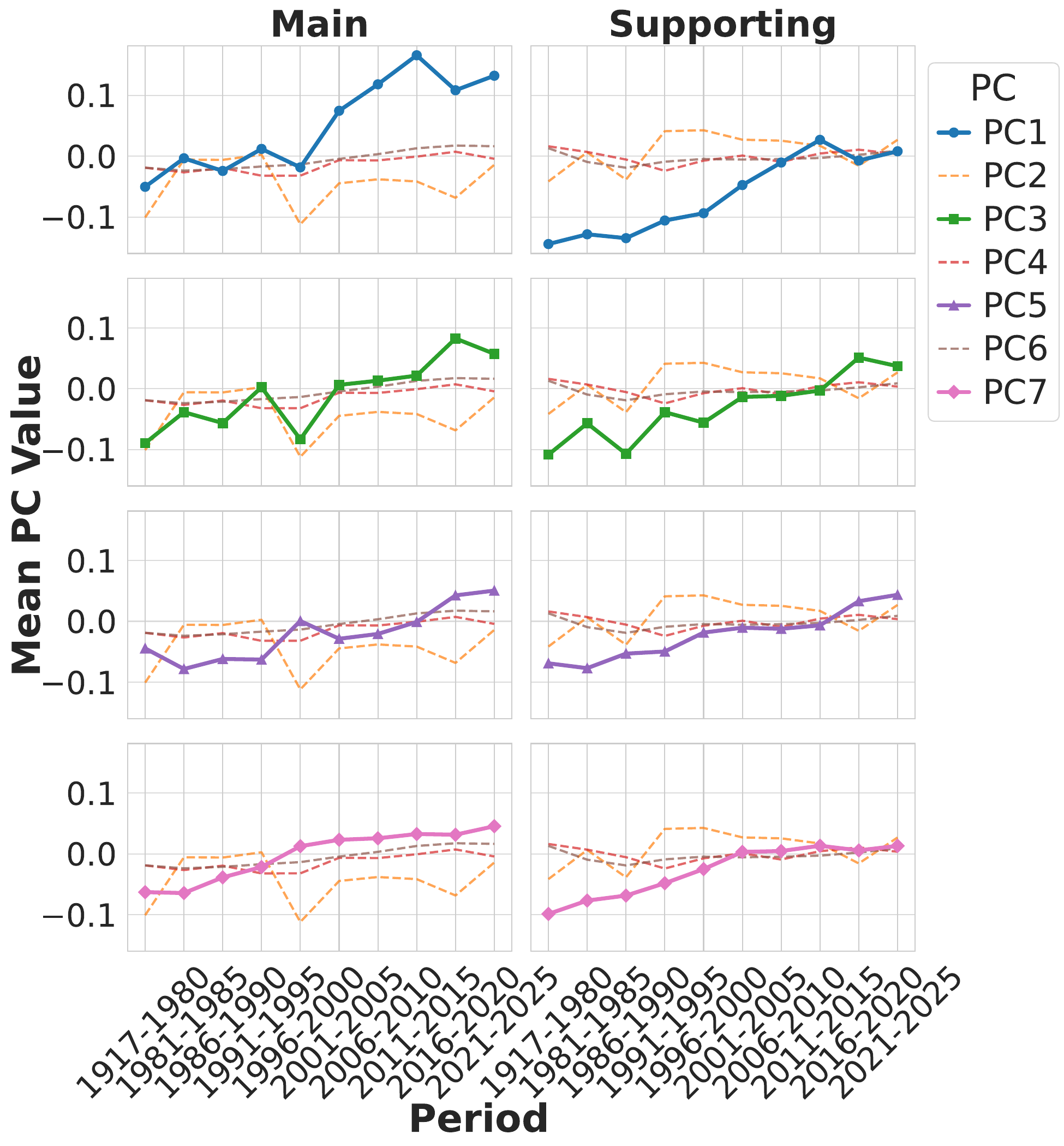}
        \caption{Mean values of the top 7 principal components of avatar embeddings across different periods. PC1, PC3, PC5, and PC7 show an increasing trend, reflecting greater use of \textbf{\textit{moe}} elements in art style designs.}
        \label{fig:avatar-pca-period}
\end{figure}

The temporal shifts in dimensions of anime avatar design are illustrated in Figure~\ref{fig:avatar-pca-period}. While most PCs are stable over time, PC1, PC3, and PC5 exhibit a marked increase over the years. As outlined earlier, PC1 reflects a shift from exaggerated masculinity toward softer, more approachable female traits. PC3 indicates a trend in body shapes toward slenderness and more exaggerated proportions, moving away from cartoonish forms. In addition, PC5 illustrates a trend toward brighter, more luminous visual designs. High values of these PCs are often associated with \textit{moe} female character design, suggesting an upward trend to incorporate more \textit{moe} elements in anime characters.

\begin{table}[t]
\centering
\begin{tabular}{lcc}
\toprule
\textbf{Model} & \textbf{F1 Score} & \textbf{Accuracy} \\
\midrule
Baseline (random) & 0.09 & 9.7\% \\
Baseline (majority) & 0.03 & 17.9\% \\
DINOv2 (frozen backbone) & 0.05 & 11.8\% \\
DINOv2 & 0.39 & 39.3\% \\
\bottomrule
\end{tabular}
\caption{Period classification performance using original character avatars across different models.}
\label{tab:model_performance}
\end{table}

\section{RQ4: Associations Between Personality Traits and Avatars}
While commonly assumed, the extent to which visual designs of characters encode personality traits lacks empirical evidence. Despite the variation in the styles of individual artists, \textbf{our analysis revealed systematic associations between the artistic rendering of avatars and personality traits}. We performed PCA on the DINOv2 embeddings of avatars and selected the top 100 PCs as image features. Using stratified sampling, we partitioned the data into 70\% training, 10\% validation, and 20\% test sets. Then Random Forest and an XGBoost models were trained to predict personality archetypes from the avatars. Both models exceeded baselines (random: F1 0.11/Acc 12.2\%; majority: F1 0.05/Acc 22.8\%), with random forest achieving 0.32 macro-F1 and 40.7--41.0\% accuracy across 20--100 PCs, and XGBoost performing similarly and peaking at 100 PCs (F1 0.34, Acc 41.2\%).
Table~\ref{tab:personality_pred} suggests that personality archetypes are highly predictable from the visual features of avatars with only 20 PCs.

The confusion matrix in Figure~\ref{fig:pc-comm-pred} in the Appendix further indicates that the personality archetypes are systematically associated with the art design. Justice Keepers, Energetic Extroverts, and Cold-hearted Antagonists are among the most predictable archetypes from the visual features, suggesting that there are highly conventional and frequently reused visual tropes for designing these character types. 

\begin{table}[t]
\centering
\begin{tabular}{lcc}
\toprule
\textbf{Model} & \textbf{F1 Score} & \textbf{Accuracy} \\
\midrule
Baseline (random) & 0.11 & 12.2\% \\
Baseline (majority) & 0.05 & 22.8\% \\
Random Forest (PC=20) & 0.32 & 40.9\% \\
Random Forest (PC=50) & 0.32 & 41.0\% \\
Random Forest (PC=100) & 0.32 & 40.7\% \\
XGBoost (PC=20) & 0.30 & 36.3\% \\
XGBoost (PC=50) & 0.34 & 39.9\% \\
XGBoost (PC=100) & 0.34 & 41.2\% \\
\bottomrule
\end{tabular}
\caption{Personality archetype classification performance using PCs of DINOv2 avatar embeddings. XGBoost substantially outperforms baselines, suggesting that visual features could indicate personality archetypes.}
\label{tab:personality_pred}
\end{table}

\section{RQ5: Modeling User Preferences}
\begin{figure}[t]
    \centering
        \includegraphics[width=0.95\linewidth]{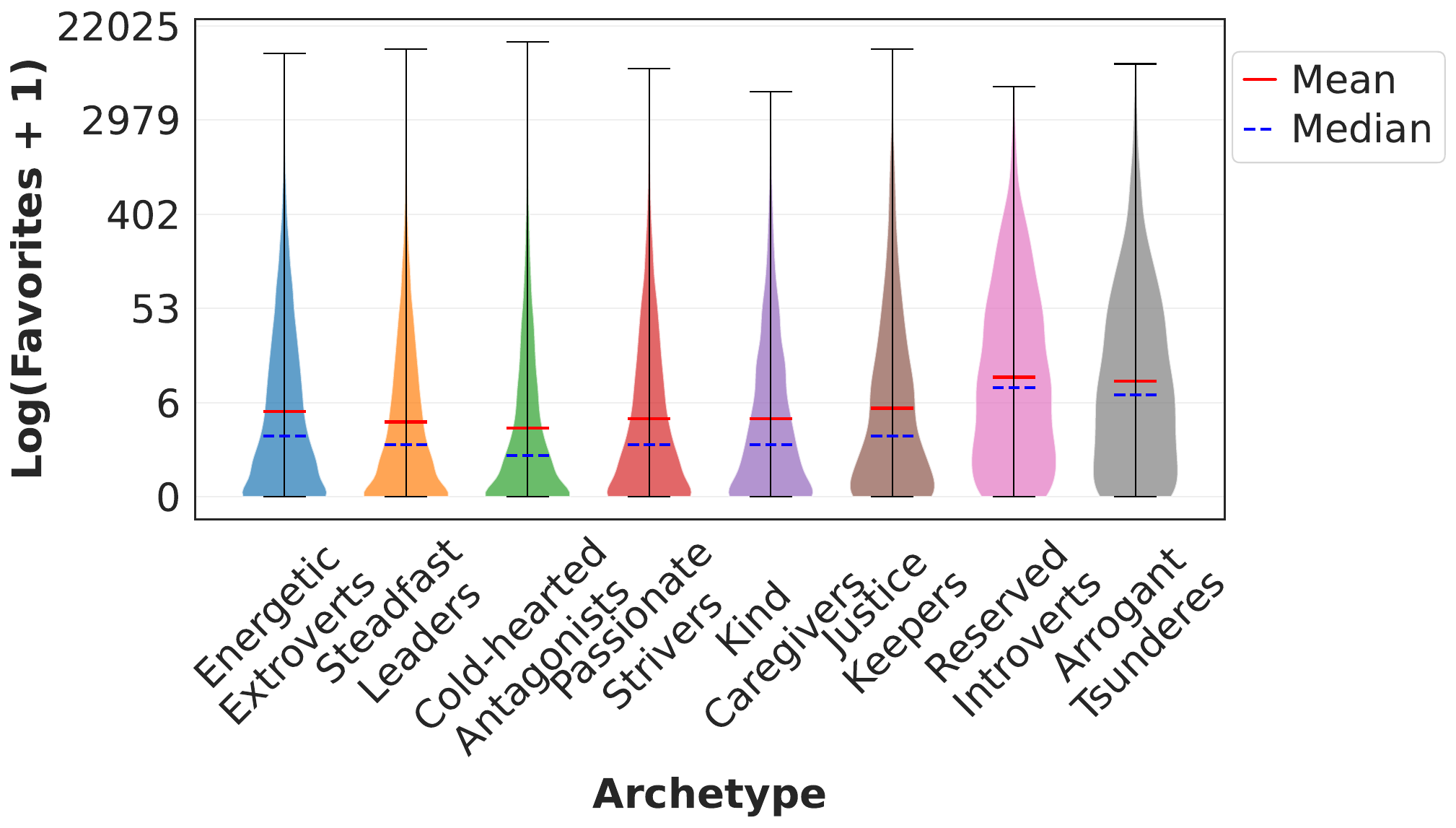}
        \caption{Distribution of log-scaled favorite counts by personality archetype. }
        \label{fig:char-pop-yearly}
\end{figure}

Prior research has suggested that anime characters' personality traits have an impact on the popularity of female characters but not male characters \cite{tang2024correlation}. Here we attempt to model user preferences using the crowd-sourced ratings on \texttt{MyAnimeList}. We examined the distribution of favorite counts across different personality archetypes on \texttt{MyAnimeList} in Figure~\ref{fig:char-pop-yearly}. Despite their small share among all archetypes, Reserved Introverts and Arrogant Tsunderes have the highest mean and median favorite counts, with wider upper tails. This may be because both archetypes are predominantly female, and traits such as being easily shy, tsundere, and contrasting personalities tend to be more popular among male audiences. While character design relies on many recurring tropes, some combinations of these tropes might make characters more popular. Motivated by this hypothesis, we design experiments to test whether character popularity is predictable from personality and visual design. 

We predicted anime character popularity (measured by the number of user favorites) using PCs derived from personality and avatar features. For each character, we used the top 20, 50, or 100 PCs as predictors and tested different feature sets: personality PCs only, avatar PCs only, both sets combined, and each of these with extra metadata (year, gender, role, anime favorites). Missing values in PCs were imputed with zero, and entries with missing metadata were dropped. Since favorites are non-negative and heavily right-skewed, we applied a $\log(1+x)$ transform on the target favorites to further stabilize variance. For prediction, Poisson regression and Histogram-based Gradient Boosted Trees (HGBT) with Poisson loss, using default model hyperparameters. The data were randomly split into train and test sets (90\%/10\%). The predictions were evaluated with mean absolute error (MAE) and mean Poisson deviance (MPD).

\begin{figure}[t]
    \centering
    \includegraphics[width=0.9\linewidth]{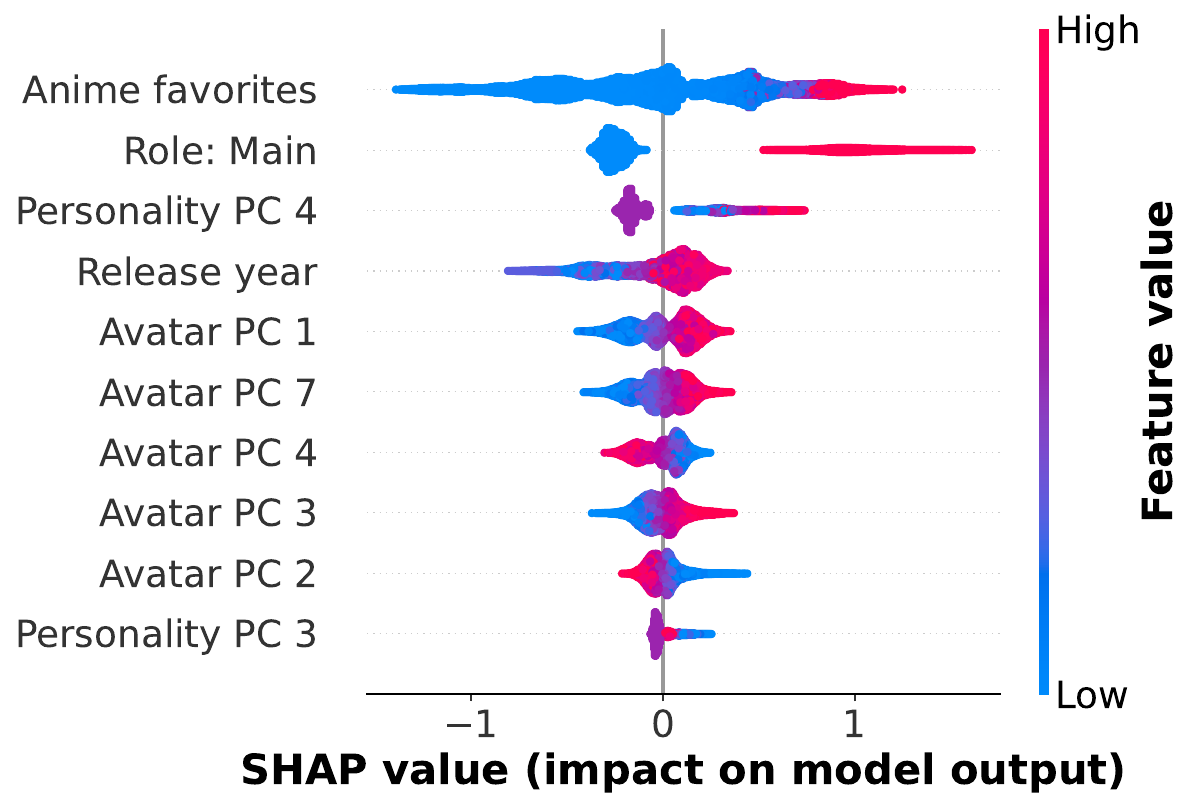}
    \caption{Beeswarm plot showing SHAP effect distributions for the top 10 features influencing character popularity prediction using HGBT (top 50 PCs from avatars and personalities plus metadata), ranked by importance.}
    \label{fig:shap}
\end{figure}

To investigate feature importance and provide interpretable insights into the best performing model (HGBT using Top 50 PCs of Avatar and Personality as well as metadata), we employed SHapley Additive exPlanations (SHAP) \cite{lundberg2017unifiedapproachinterpretingmodel} to quantify the feature contributions, as shown in Figure~\ref{fig:shap}. Metadata features of anime favorites, roles, and release year demonstrate high importance. \textbf{This aligns with expected patterns that characters from more popular anime series and main characters from more recent years naturally gain more audience attention and favorability}. In addition to metadata, the character design is also associated with audience preference. Specifically, \textbf{visual design plays a much more crucial role than personality features}, likely because anime characters typically follow a fixed set of personality archetypes with limited variation. As illustrated in the figure, PC1, PC7, PC4, PC3, and PC2 show significant contrasts in SHAP values across different feature values. Based on our previous interpretation of these principal components of avatar visual features, characters are generally more favored when they exhibit softer, rounder, and more feminine facial structures (PC1), probably catering to the male preference for female characters with neotenous, \textbf{\textit{moe}}-style features \cite{matsui_moe-phobia_2022}. Besides, those who wear masks, armors, or helmets (PC7) also demonstrate higher popularity, as this delicate and elaborate design of characters makes them more visually intriguing and cool to audiences. Artistic designs featuring proportionally slender bodies, oval faces (PC3), long hair, and goddess-like aesthetics (PC4) are also highly favored, potentially reflecting curvaceous female features intended to appeal to male viewers \cite{reysen_examination_2017, sym12081261}.

%BEGIN DIFADD
\begin{table}[t]
\centering
\begin{tabular}{lllcc}
\toprule
\textbf{Features}  & \textbf{PCs}  & \textbf{Model}  & \textbf{MAE}  & \textbf{MPD} \\
\midrule
Baseline (mean)  & --  & --  & 1.190  & 2.166 \\ 
\midrule
Both PCs + Metadata  & 20  & Poisson  & 0.897  & 1.416 \\ 
Both PCs + Metadata  & 20  & HGBT  & 0.669  & 0.946 \\ 
Both PCs + Metadata  & 50  & Poisson  & 0.886  & 1.385 \\ 
Both PCs + Metadata  & 50  & HGBT  & \textbf{0.668}  & \textbf{0.945} \\ 
\midrule
Avatar PCs only  & 20  & Poisson  & 1.022  & 1.725 \\ 
Avatar PCs only  & 20  & HGBT  & 1.015  & 1.693 \\ 
Avatar PCs only  & 50  & Poisson  & 1.001  & 1.666 \\ 
Avatar PCs only  & 50  & HGBT  & 1.008  & 1.673 \\ 
\midrule
Personality PCs only  & 20  & Poisson  & 1.119  & 1.997 \\ 
Personality PCs only  & 20  & HGBT  & 0.972  & 1.590 \\ 
Personality PCs only  & 50  & Poisson  & 1.110  & 1.979 \\ 
Personality PCs only  & 50  & HGBT  & 0.970  & 1.585 \\ 
\midrule
Metadata only  & --  & Poisson  & 0.994  & 1.651 \\ 
Metadata only  & --  & HGBT  & 0.779  & 1.204 \\ 
\bottomrule
\end{tabular}
\caption{Results of predicting log-transformed favorite counts using avatar PCs, personality PCs, and metadata (anime favorites, gender, role, year). Lower MAE and MPD suggest better performance.}
\label{tab:fav_pred_results}
\end{table}
%END DIFADD

\section{Discussions}
Here we would like to discuss in depth the implications of these empirical findings.

\subsubsection{Anime for a Maturing Audience}
Our analysis indicates a systematic shift in anime's target audience, moving from one primarily for children to a more maturing audience of teenagers and young adults (the \textit{Shonen} population) in recent decades (\textbf{RQ1}). Such a shift has also been manifested in multiple aspects. The narrative has become more complex, with more supporting characters and more psychologically complex main characters. For example, the decline of manipulative anti-heroes signifies a deviation from the simple plots driven by the struggle against evil. The rise of complex and flawed protagonists like emotional introverts also suggests a move towards more relatable characters with intense emotional conflicts. Similarly, in terms of visual design, the characters are also designed with more idealized and exaggerated body ratios (Avatar PC3) and more sexualized designs (Avatar PC3 and PC5). Yet anime with more explicit content for older adults is in decline, after a brief peak in the 2000s. Contemporary anime is mostly defined by the \textit{Shonen} anime \cite{drummond2010boys}. 

\subsubsection{\textit{Moe}-ification of Anime Characters}
A prominent trend emerging from the data is the tendency of \textit{moe}-ification (\textbf{RQ2 \& 3}). In terms of personality traits, two archetypes, Reserved Introverts and Arrogant Tsunderes, have witnessed a noticeable rise in recent decades. These two archetypes, dominated by female characters, rank among the most popular archetypes. The appeal of introverts draws viewers through quiet charm and vulnerability, while tsundere characters display contrasting traits of arrogance and hidden warmth, an attractive feature known as \texttt{gap-moe}. The visual features that have undergone the most significant changes are all features related to \textit{moe} or sexualized female traits (Avatar PC1 and PC3). The \textit{moe}-ified visual features are also strongly predictive of audience preference on \texttt{MyAnimeList}, where the audience is mostly outside Japan and consists of 75\% males. This suggests a more universal appeal of \textit{moe} design among the male audience that transcends national boundaries. 

Commercially, this has been known to be a deliberate choice of the designers driven by economic gains \cite{galbraith2014moe}. In the media-mix strategy \cite{steinberg2012anime,yatron202230}, anime is only a part of the commercialization scheme, and popular characters created in the anime will be further turned into various merchandise. Through commodifying affection, the anime studios can push for maximum profits from the primary consumer base, the Otaku community \cite{mcveigh1996commodifying}, who would form a strong attachment to \textit{moe} characters and are willing to consume character-related spin-offs. Yet, new from prior studies, our analysis quantitatively reveals that the \textit{moe}-ification trend becomes increasingly prominent after 2000, especially among main characters, despite the international expansion of audience \cite{noh2024global}. This reflects an ongoing structural shift in the anime industry, a shift towards less diversity in content design and more pandering to the male Otaku community. It may create a self-reinforcing feedback loop, as the prevailing design conditions the preferences of the expanding market worldwide to align with such tastes, thereby shaping audience preferences across diverse cultural contexts, as well as global animation production practices.

Despite the commercial success, it is also hard to omit the gender biases and female objectification in such \textit{moe} character design. This is coupled by the fact that female characters are more likely to be associated with stereotypical personality traits like naivety, shyness, supportiveness, and emotional volatility. Such highly conventional and objectifying portrayals of femininity, with some infantilized as immature, innocent, and docile (Avatar PC1) and others sexualized as highly sensual and curvaceous (Avatar PC3 and PC5), have long been criticized \cite{ting2019gender,lunning2011under,reysen_examination_2017}. Sustained exposure to stereotypical gender representations may contribute to adverse effects on self-image and mental health, as well as the normalization of sexism, especially among young female audiences \cite{ward2020media,ijerph20105770,cho2025sexual}. However, the \textit{moe} design cannot always be reduced to male-pandering, as some argue that the \textit{moe} design can also be empowering, subverting societal norms, and asserting queer identities \cite{akgun2020mythology,yatron202230, saito2014magic}.

\subsubsection{Creativity in Cultural Products}
Japanese anime distinguishes itself from comics, cartoons, and other cultural products through the recognizable `anime-esque' traits from a database of conventionalized models \cite{suan2017anime,ruh2014conceptualizing,azuma2009otaku}. Therefore, anime creativity is often confined to local variations of the global uniformity and repetition of anime-esque elements \cite{suan2017anime}. In our analysis, the relative proportion of different personality archetypes are generally stable over time, reflecting an embrace of such uniformity. We also map out these local variations of anime character design and how they shift over time with large-scale computational methods. Only 20 PCs are sufficient to explain 57.88\% of variance of character avatars. 

Personality archetypes are often predictable from visual features, especially for certain flat archetypes (\textbf{RQ4}). This aligns well with empirical evidence from existing research that facial features are a significant factor in personality perception among the anime audience \cite{ze2025PERCEPTIONOA}, and that viewers develop learned expectations about how specific visual features correspond to personality \cite{chen2015relp, NAKASHIMA20231776}. Yet our findings indicate that the audience seems to prefer personality archetypes that are less predictable from their visual designs, or less conventionalized characters.

Our modeling of audience preferences demonstrates that audience preferences are driven by the integration of artistic styles and personality traits. Yet the visual features play a more dominant role than personality traits, even after controlling for the meta-information like role and anime popularity (\textbf{RQ5}). Feminine traits such as \textit{moe}-style faces, long stylish hair, and slender proportional bodies are associated with higher popularity. Characters with elaborate mechanical designs, like armor and giant humanoid robots (or \textit{Mecha}), also contribute to popularity. Despite the large variation in character design, audience preference is consistently linked to certain aesthetic choices from the anime database \cite{ruh2014conceptualizing,azuma2009otaku}. We also found these two design elements, \textit{moe} and \textit{Mecha}, have exhibited a growing trend in the visual design. Our findings also align with the observation that anime creativity is more on mixing and matching of conventional anime-esque elements \cite{suan2021anime}, a modular production style that ensures these artistic designs retain their appeal regardless of genre boundaries, the expanding creator base, or changing market trends.

\section{Conclusions}
In this study, we analyze the character design in Japanese anime with large-scale multimodal data from an anime review site. By combining textual, visual, and production features of anime characters with online popularity traces, we are able to reveal the historical trend of anime as a genre. Our results suggest that anime designers do utilize recurring tropes in constructing character personalities and designing their visual representations. The design has been undergoing \textit{moe}-ification since the 2000s, suggesting increased gendered objectification of females in the modern anime production industry. This phenomenon might be worth further examination to understand the causes and its influence on the audience. Overall, our research demonstrates that large-scale analysis of online cultural artifacts can provide insights into cultural analytics.

\section{Limitations}
This study is still limited in several ways. First, the data were gathered from \texttt{MyAnimeList}, which could potentially be biased towards English cultures and values. Secondly, due to the crowd-sourced nature of the site, the representations of anime in \texttt{MyAnimeList} are biased towards popular and contemporary works. Thirdly, due to the limitation in data, we were not able to analyze the vocal design, which is also critical to the appeal of anime characters \cite{starr2015sweet}. In addition, the incompleteness of genre and theme metadata precluded analyses across different genres. Finally, some findings may be affected by mediating factors that are unobserved in the curated dataset (e.g., cultural context, content rating mechanisms, and temporal factors), and should be interpreted with caution. Future research could potentially mitigate these biases with richer multimodal data from more diverse cultural communities.

\section{Acknowledgments}
We thank the reviewers and the area chair for their helpful comments. This research was enabled in part through the computational resources provided by Advanced Research Computing at the University of British Columbia and the Digital Research Alliance of Canada. The research activities were also supported by the NSERC Discovery Grant and the CFI JELF Grant awarded to JZ.

\bibliography{aaai2026}

\section{Paper Checklist}

\begin{enumerate}

\item For most authors...
\begin{enumerate}
    \item  Would answering this research question advance science without violating social contracts, such as violating privacy norms, perpetuating unfair profiling, exacerbating the socio-economic divide, or implying disrespect to societies or cultures?
    \answerYes{Yes. Our study analyzes anime characters using only public data, acknowledges cultural biases, and involves no privacy or ethical violations.}
  \item Do your main claims in the abstract and introduction accurately reflect the paper's contributions and scope?
    \answerYes{Yes.}
   \item Do you clarify how the proposed methodological approach is appropriate for the claims made? 
    \answerYes{Yes. See the \textbf{Method} section.}
   \item Do you clarify what are possible artifacts in the data used, given population-specific distributions?
    \answerYes{Yes. We discuss the potential biases from \texttt{MyAnimeList}'s English-oriented community and how representations of anime are skewed toward popular and contemporary works in the \textbf{Limitation} section.}
  \item Did you describe the limitations of your work?
    \answerYes{Yes. See the \textbf{Limitations} section.}
  \item Did you discuss any potential negative societal impacts of your work?
    \answerNo{No. Our research analyzes fictional characters using public data and does not pose societal risks.}
      \item Did you discuss any potential misuse of your work?
    \answerNo{No. Our findings are limited on fictional characters and do not enable harmful applications or misues.}
    \item Did you describe steps taken to prevent or mitigate potential negative outcomes of the research, such as data and model documentation, data anonymization, responsible release, access control, and the reproducibility of findings?
    \answerNA{NA.}
  \item Have you read the ethics review guidelines and ensured that your paper conforms to them?
    \answerYes{Yes.}
\end{enumerate}

\item Additionally, if your study involves hypotheses testing...
\begin{enumerate}
  \item Did you clearly state the assumptions underlying all theoretical results?
    \answerNA{NA.}
  \item Have you provided justifications for all theoretical results?
    \answerNA{NA.}
  \item Did you discuss competing hypotheses or theories that might challenge or complement your theoretical results?
    \answerNA{NA.}
  \item Have you considered alternative mechanisms or explanations that might account for the same outcomes observed in your study?
    \answerNA{NA.}
  \item Did you address potential biases or limitations in your theoretical framework?
    \answerNA{NA.}
  \item Have you related your theoretical results to the existing literature in social science?
    \answerNA{NA.}
  \item Did you discuss the implications of your theoretical results for policy, practice, or further research in the social science domain?
    \answerNA{NA.}
\end{enumerate}

\item Additionally, if you are including theoretical proofs...
\begin{enumerate}
  \item Did you state the full set of assumptions of all theoretical results?
    \answerNA{NA.}
	\item Did you include complete proofs of all theoretical results?
    \answerNA{NA.}
\end{enumerate}

\item Additionally, if you ran machine learning experiments...
\begin{enumerate}
  \item Did you include the code, data, and instructions needed to reproduce the main experimental results (either in the supplemental material or as a URL)?
    \answerYes{Yes.}
  \item Did you specify all the training details (e.g., data splits, hyperparameters, how they were chosen)?
    \answerYes{Yes. The data splits are listed in the main text, and the fine-tuning hyperparameters for DINOv2 are summarized in Table~\ref{table:dinov2_finetuning_hyperparams}. More details can be found in the code provided in the GitHub repository.}
     \item Did you report error bars (e.g., with respect to the random seed after running experiments multiple times)?
    \answerNo{No.}
	\item Did you include the total amount of compute and the type of resources used (e.g., type of GPUs, internal cluster, or cloud provider)?
    \answerYes{Yes, specified in the \textbf{Resources} section in the appendix.}
     \item Do you justify how the proposed evaluation is sufficient and appropriate to the claims made? 
    \answerYes{Yes. As specified in the \textbf{Method} section, our evaluations are tailored to the research questions and directly support our claims.}
     \item Do you discuss what is ``the cost`` of misclassification and fault (in)tolerance?
    \answerNA{NA.}
  
\end{enumerate}

\item Additionally, if you are using existing assets (e.g., code, data, models) or curating/releasing new assets, \textbf{without compromising anonymity}...
\begin{enumerate}
  \item If your work uses existing assets, did you cite the creators?
    \answerYes{Yes. We have cited \texttt{MyAnimeList} for data, as well as pretrained models and relevant prior works used for feature extraction and analysis.}
  \item Did you mention the license of the assets?
    \answerYes{Yes, details about the licenses for the models (\texttt{Camie Tagger}, \texttt{dinov2-base}, \texttt{Qwen3-32B-FP8}, and \texttt{Qwen3-Embedding-0.6B}) are provided in the \textbf{Licenses} section in the appendix.}
  \item Did you include any new assets in the supplemental material or as a URL?
    \answerNo{No. We do not release any new assets.}
  \item Did you discuss whether and how consent was obtained from people whose data you're using/curating?
    \answerNA{NA.}
  \item Did you discuss whether the data you are using/curating contains personally identifiable information or offensive content?
    \answerNo{No. All data come from publicly available sources on \texttt{MyAnimeList} and do not contain personal information or offensive content.}
\item If you are curating or releasing new datasets, did you discuss how you intend to make your datasets FAIR (see \citet{fair})?
\answerNA{NA.}
\item If you are curating or releasing new datasets, did you create a Datasheet for the Dataset? 
\answerNA{NA.}
\end{enumerate}

\item Additionally, if you used crowdsourcing or conducted research with human subjects, \textbf{without compromising anonymity}...
\begin{enumerate}
  \item Did you include the full text of instructions given to participants and screenshots?
    \answerYes{Yes. The full human annotation instructions and an interface screenshot are provided in the Appendix.}
  \item Did you describe any potential participant risks, with mentions of Institutional Review Board (IRB) approvals?
    \answerNA{NA.}
  \item Did you include the estimated hourly wage paid to participants and the total amount spent on participant compensation?
    \answerNo{No. Since the annotation task was performed by two of the authors, no participant compensation was required.}
   \item Did you discuss how data is stored, shared, and deidentified?
   \answerYes{Yes. Details are provided in the Appendix.}
\end{enumerate}

\end{enumerate}

\appendix

\section{Licenses} \label{appdx:licenses}

Pretrained models used in this study include \texttt{Camie Tagger}, which is distributed under the GNU General Public License v3.0, as well as \texttt{DINOv2-base}, \texttt{Qwen3-32B-FP8}, and \texttt{Qwen3-Embedding-0.6B}, all of which are released under the Apache License 2.0.

\section{Resources} \label{appdx:resources}

All model training and experiments were conducted on an institutional high-performance computing cluster using the 20 GB H100 GPU allocation, providing sufficient capacity for both LLM annotation and Vision Transformer fine-tuning. Other computations, such as community detection and principal component analysis, were executed using up to 16 CPUs and 32GB of RAM on the cluster.

\section{Data Storage and Privacy} \label{appdx:data-privacy}

All data were stored on the cluster and were only accessible to the authors. The data contains no personally identifiable information, and no de-identification was required.

\section{Disclosure on the Use of Generative AI} \label{appdx:ai-use}

We confirm that all text in the manuscript was written by the authors, with no original intellectual contributions generated by AI. We solely utilized AI tools for assisting with research code, grammar and spelling checks, as well as translating obscure vocabulary and rephrasing.

\section{LLM Prompts} \label{appdx:prompt}

\begin{tcolorbox}[colback=gray!10, colframe=gray!80!black, title=LLM Prompt for character personality trait extraction]

You are an expert Japanese anime character analyst. Based on the character's biography, provide a structured JSON description.

Describe personality traits using a small set of Japanese anime-style keywords in two languages:
\begin{list}{$-$}{\leftmargin=1.8em \labelwidth=1em \labelsep=0.4em}
  \item Japanese: short descriptors commonly used in anime character analysis.
  \item English: natural English equivalents in the same order.
\end{list}

Gender: M / F / Unknown.\\
Character JSON: the character JSON filename, including the .json extension.\\

Instructions:
\begin{list}{$-$}{\leftmargin=1.8em \labelwidth=1em \labelsep=0.4em}
  \item Output only the JSON object. Do not include any explanation, commentary, or extra text.
  \item If the biography contains clear details to infer personality traits, extract them normally.
  \item If the biography does not provide enough information to infer personality traits, then:
    \begin{list}{$\bullet$}{\leftmargin=1.5em \labelwidth=0.8em \labelsep=0.4em}
      \item Set “personality\_keywords” to \{“Japanese”: [], “English”: []\}.
    \end{list}
  \item For gender inference:
    \begin{list}{$\bullet$}{\leftmargin=1.5em \labelwidth=0.8em \labelsep=0.4em}
      \item Set “gender” to M or F only if clearly stated or implied (e.g., pronouns).
      \item Otherwise, set “gender” to “Unknown”.
    \end{list}
  \item Do not guess or assume any information not supported by the biography.
\end{list}

Output Structure:
\begin{verbatim}
{
  "character_name": "",
  "personality_keywords": {
    "Japanese": [],
    "English": []
  },
  "gender": "",
  "character_json": ""
}
\end{verbatim}

Now analyze this character:\\
Character name: \{character\_name\}\\
Biography: \{biography\}\\
Character JSON: \{character\_json\}

\end{tcolorbox}

\section{Annotation Interface Screenshot} \label{appdx:annotation-interface}

\begin{figure}[H]
    \centering
    \includegraphics[width=\linewidth]{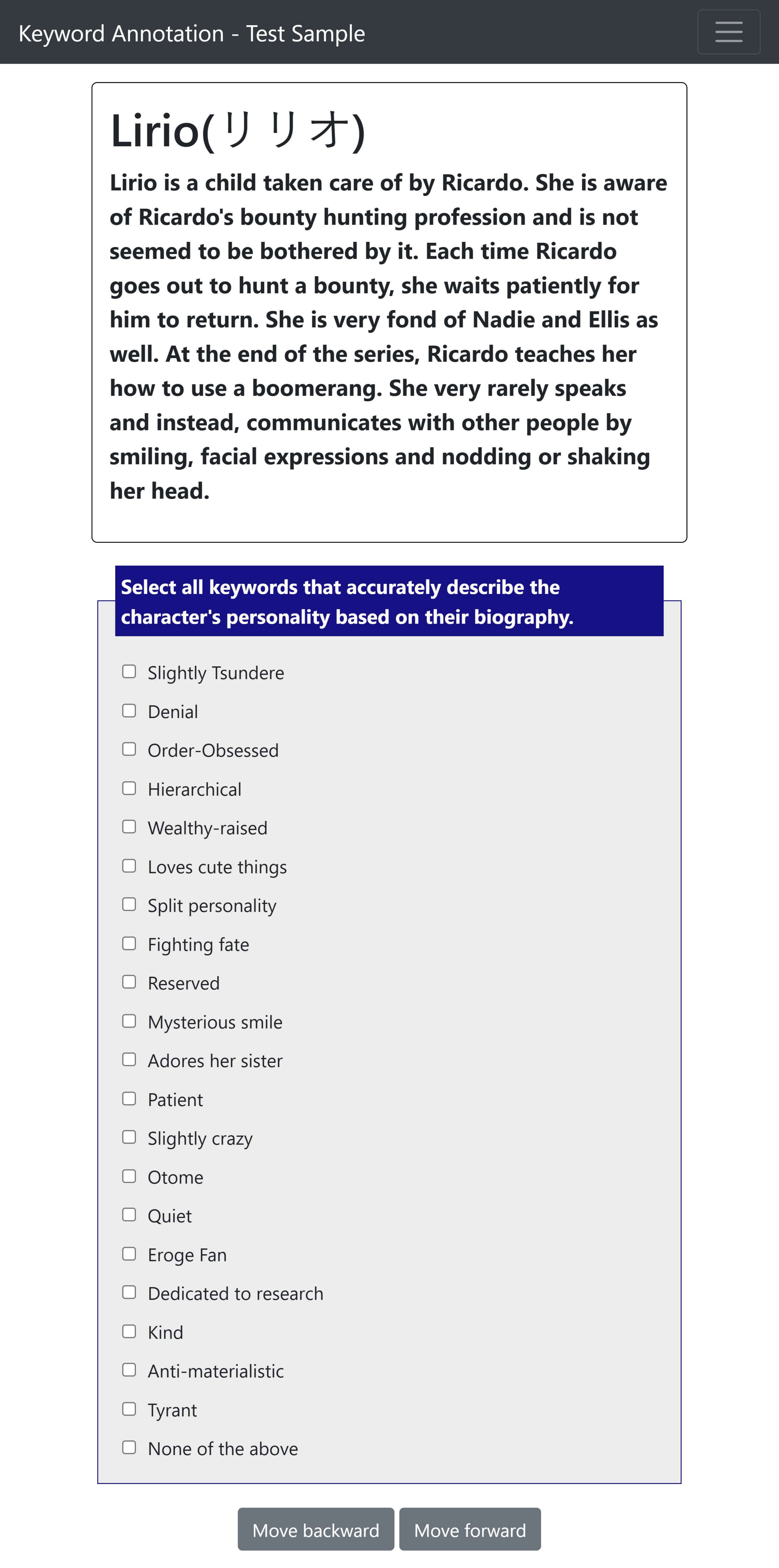}
    \caption{Screenshot of the Potato annotation interface used in the model validation task. The example shown is for illustration only and was not part of the annotated set.}
    \label{fig:interface-screenshot}
\end{figure}

\section{Annotation Guidelines} \label{appdx:annotation-guidelines}

\begin{tcolorbox}[colback=gray!10, colframe=gray!80!black, title=Annotation Guideline for Model Validation]

The goal of this task is to assess whether personality keywords extracted by the \texttt{Qwen3-32B-FP8} model are accurate based on character biographies.

For each character, you will be shown:
\begin{list}{$-$}{\leftmargin=1.2em \labelwidth=1em \labelsep=0.4em}
  \item A character biography.
  \item A shuffled list of candidate personality keywords.
\end{list}

Instructions:
\begin{list}{$-$}{\leftmargin=1.2em \labelwidth=1em \labelsep=0.4em}
  \item Read the biography carefully.
  \item Select all keywords that are clearly supported by the biography.
  \item If none apply, select ``None of the above''.
\end{list}

\end{tcolorbox}

\section{Inter-Annotator Agreement Metrics} \label{appdx:inter-annotator-agreement}

\begin{table}[H]
    \centering
    \begin{tabular}{cccc}
    \toprule
    \textbf{Precision} & \textbf{Recall} & \textbf{F1} & \textbf{Jaccard} \\
    \midrule
    0.649 & 0.770 & 0.705 & 0.551 \\
    \bottomrule
    \end{tabular}
    \caption{Inter-annotator agreement metrics between the two independent human annotators for 50 random samples. To compute directional metrics, Annotator 1 is treated as the reference and Annotator 2 as the prediction: Precision (fraction of Annotator 2's selected keywords chosen by Annotator 1), Recall (fraction of Annotator 1's selected keywords chosen by Annotator 2), F1-score, and Jaccard similarity (intersection over union of selected keyword sets). Given the great complexity of the multi-label annotation task, these metrics indicate a high degree of inter-annotator agreement.}
    \label{tab:annotator_agreement}
\end{table}

\section{Yearly Proportion of Anime Sources} \label{appdx:source-per-year}

\begin{figure}[H]
    \centering
    \includegraphics[width=\linewidth]{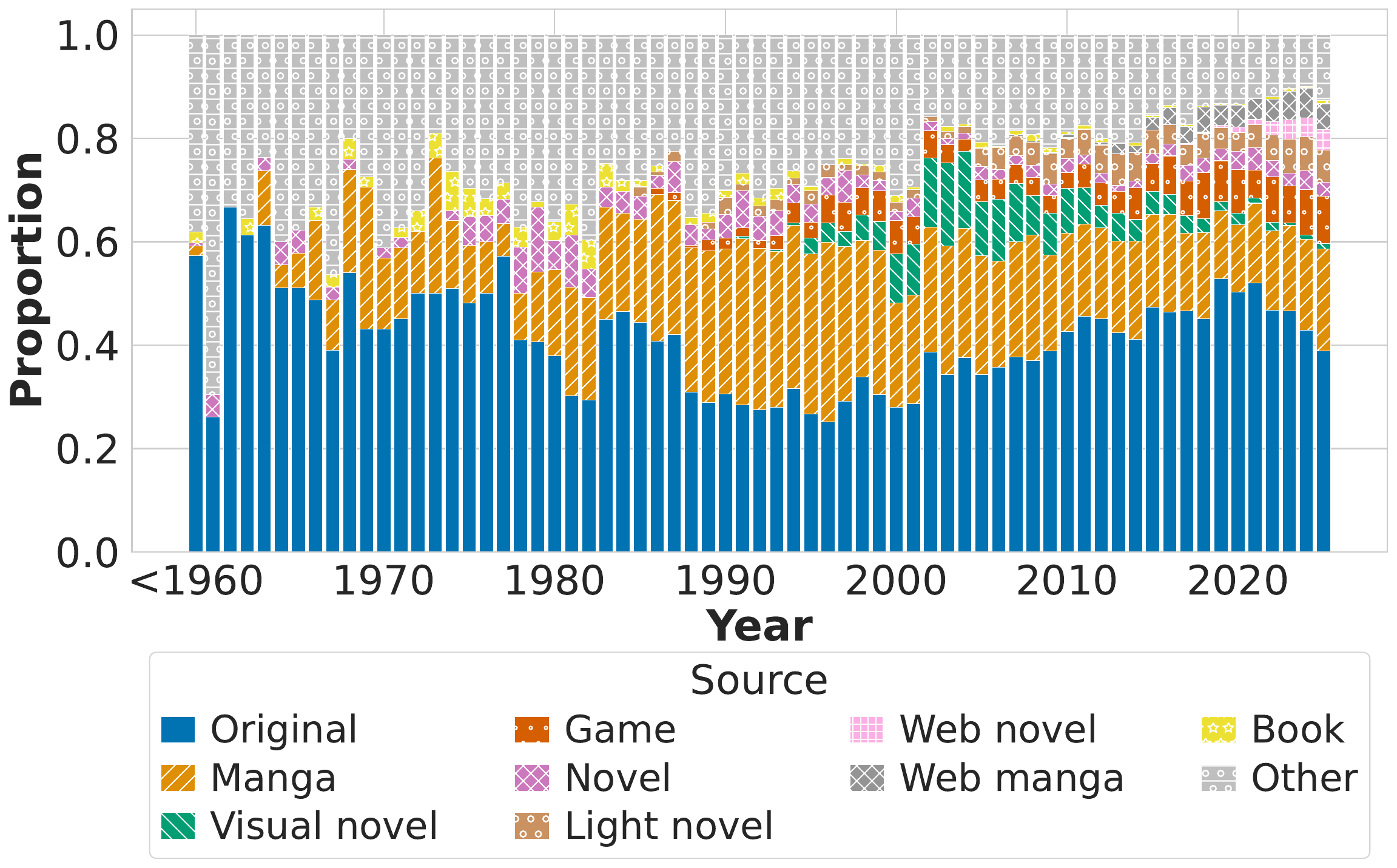}
    \caption{Proportion of source types by year. The sources of anime creation have been diversified in recent years. In addition to original creation and manga adaptation, new anime series are also adapted from games, light novels, web novels, and visual novels.}
    \label{fig:source-year}
\end{figure}

\section{Character Archetype Community Detection} \label{appdx:comm-detection}

\begin{figure}[H]
    \centering
    \includegraphics[width=0.48\textwidth]{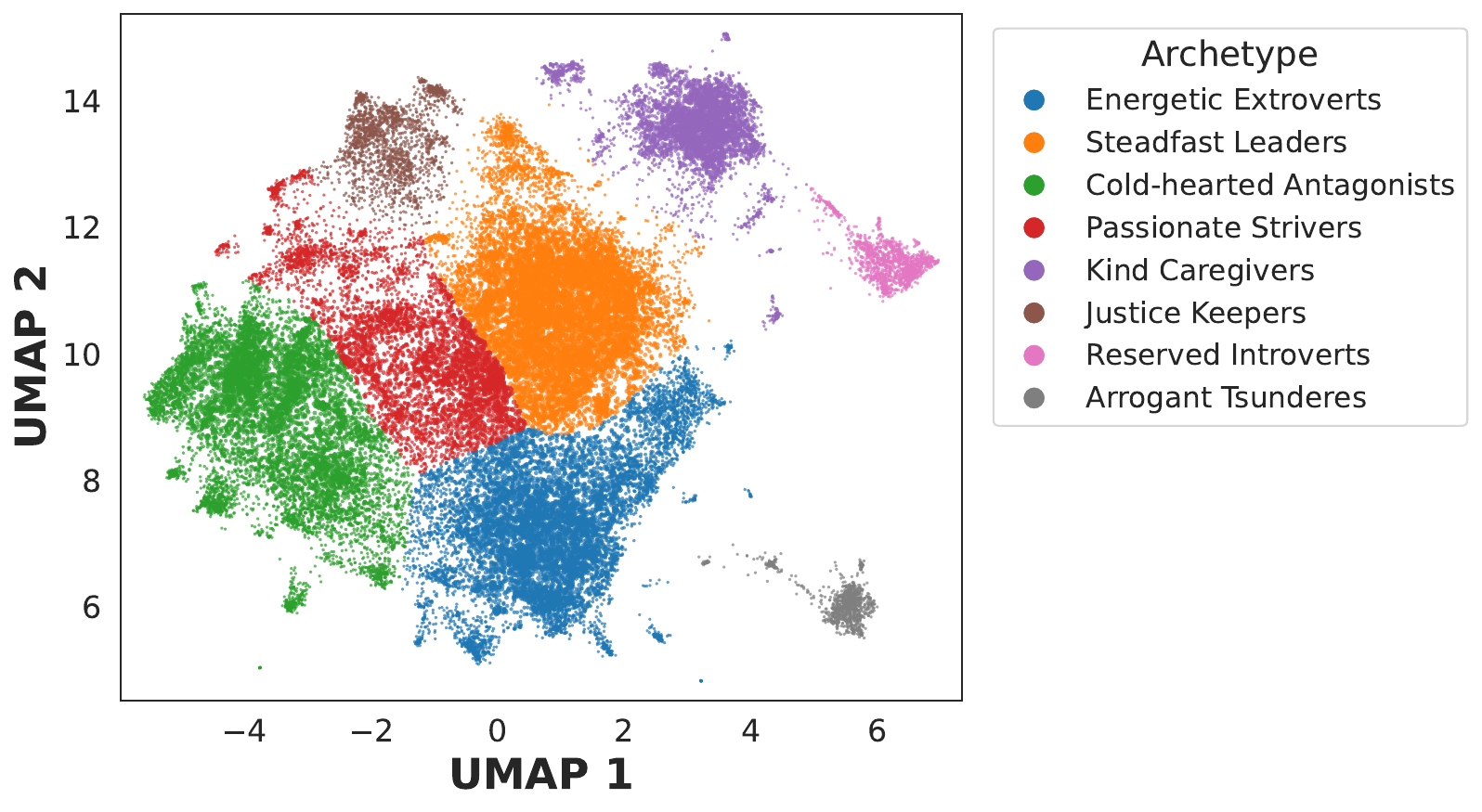}
    \caption{Character archetype communities obtained by applying Leiden algorithm to a distance‑thresholded (5th percentile) graph over the 2D UMAP coordinates; marker size reflects node degree centrality. The modularity score achieves 0.72, suggesting a strong community structure with highly separable clusters.}
    \label{fig:comm-detection}
\end{figure}

\section{Character Archetypes by Anime Rating} \label{appdx:archetype-rating}

\begin{figure}[H]
    \centering
        \includegraphics[width=0.42\textwidth]{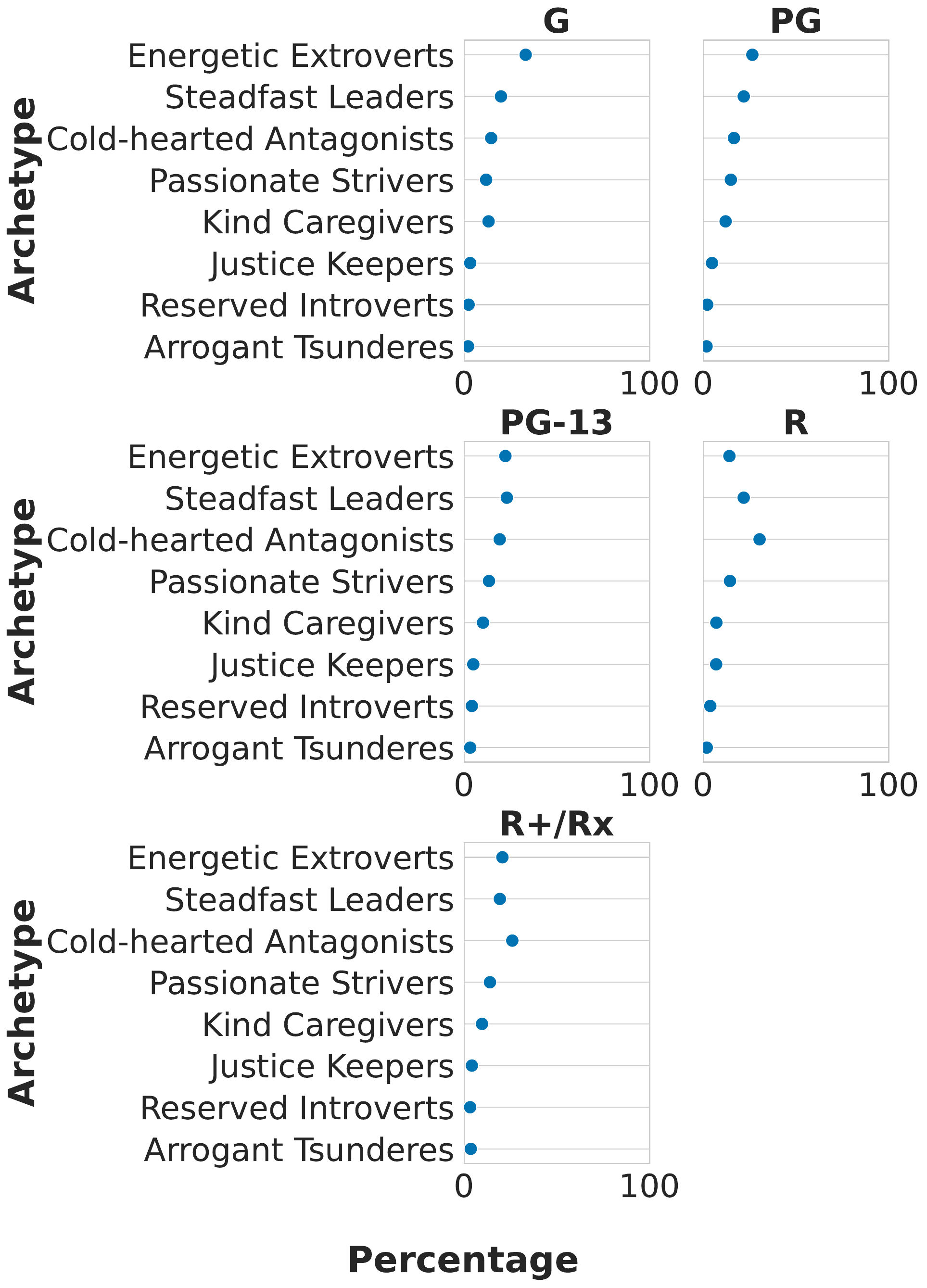}
        \caption{Archetype share across rating categories. More mature anime features darker antagonistic archetypes.}
        \label{fig:community-vs-rating}
\end{figure}

\section{Character Archetypes by Anime Source} \label{appdx:archetype-source}

\begin{figure}[H]
    \centering
        \includegraphics[width=0.42\textwidth]{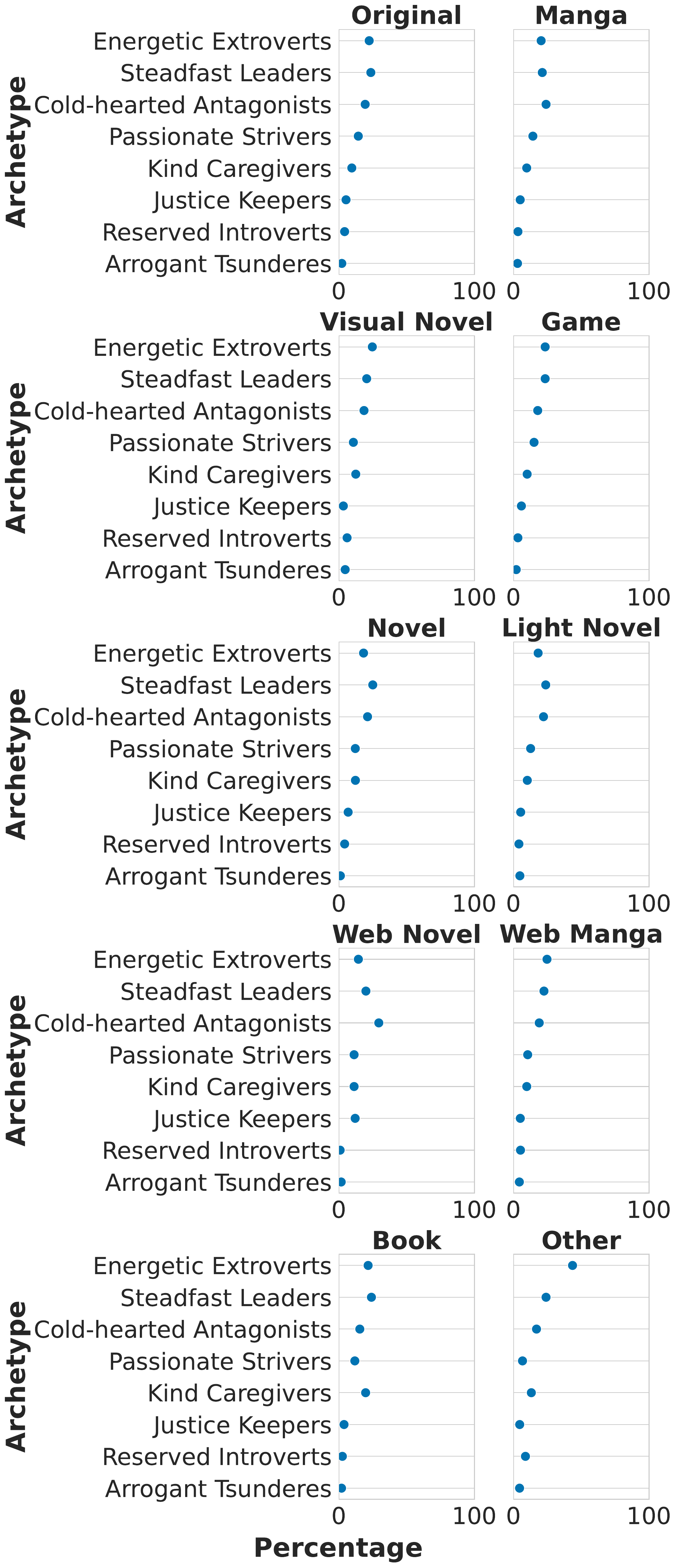}
        \caption{Archetype share across anime sources. Most sources maintain a diverse mix of character types. Web novel adaptations feature more Cold-hearted Antagonists and fewer Energetic Extroverts, reflecting darker narratives, while visual novel and web manga adaptations have more Energetic Extroverts, highlighting more optimistic themes.}
        \label{fig:community-vs-source}
\end{figure}

\section{Top/Bottom Characters in Top 10 Avatar PCs}

\begin{figure}[H]
  \centering
  \includegraphics[width=0.48\textwidth]{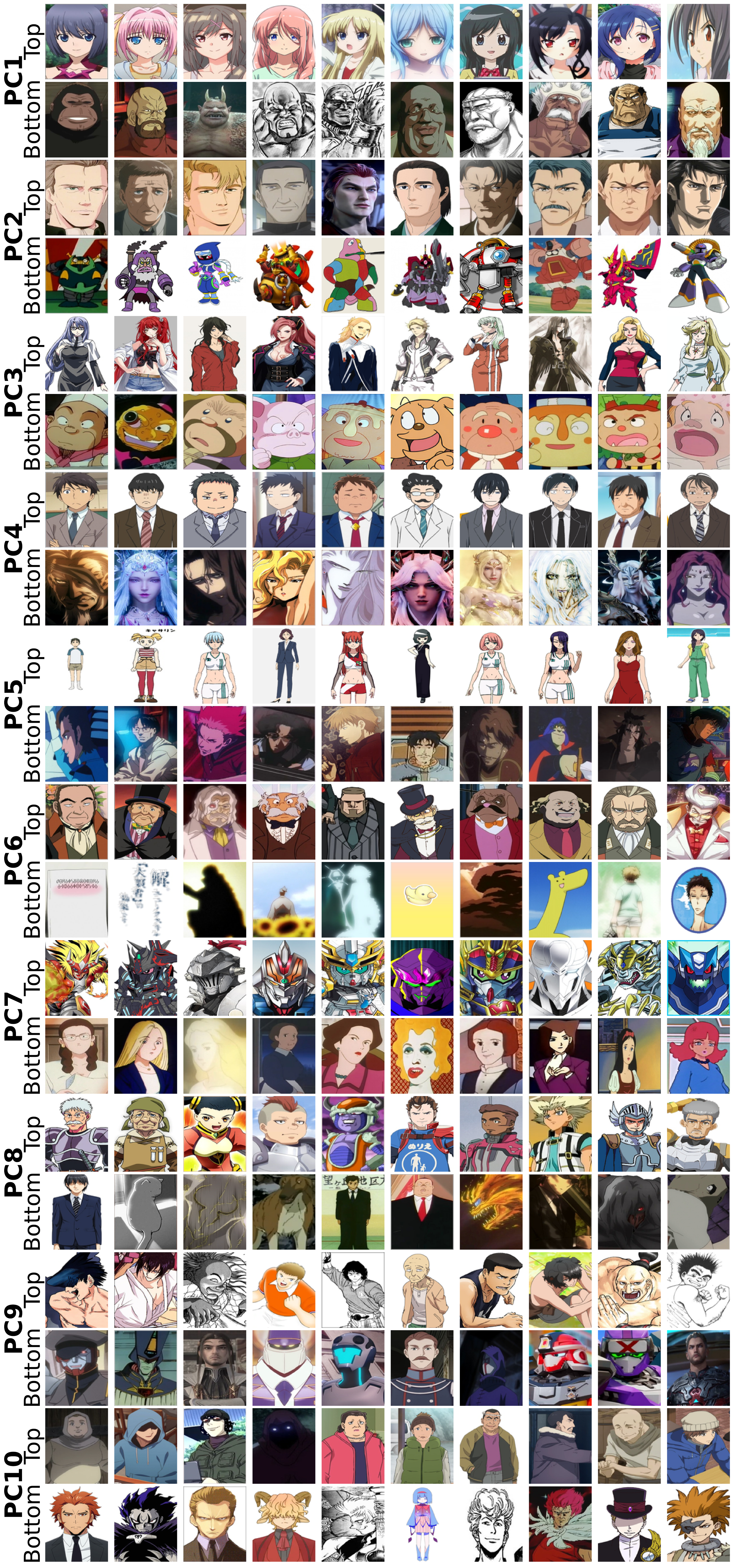}
  \caption{Top 10 and bottom 10 character examples for the first 10 principal components of avatar embeddings.}
  \label{fig:all-pc-image-example}
\end{figure}

\section{Confusion Matrix for Temporal Period Prediction}

\begin{figure}[H]
    \centering
        \includegraphics[width=\linewidth]{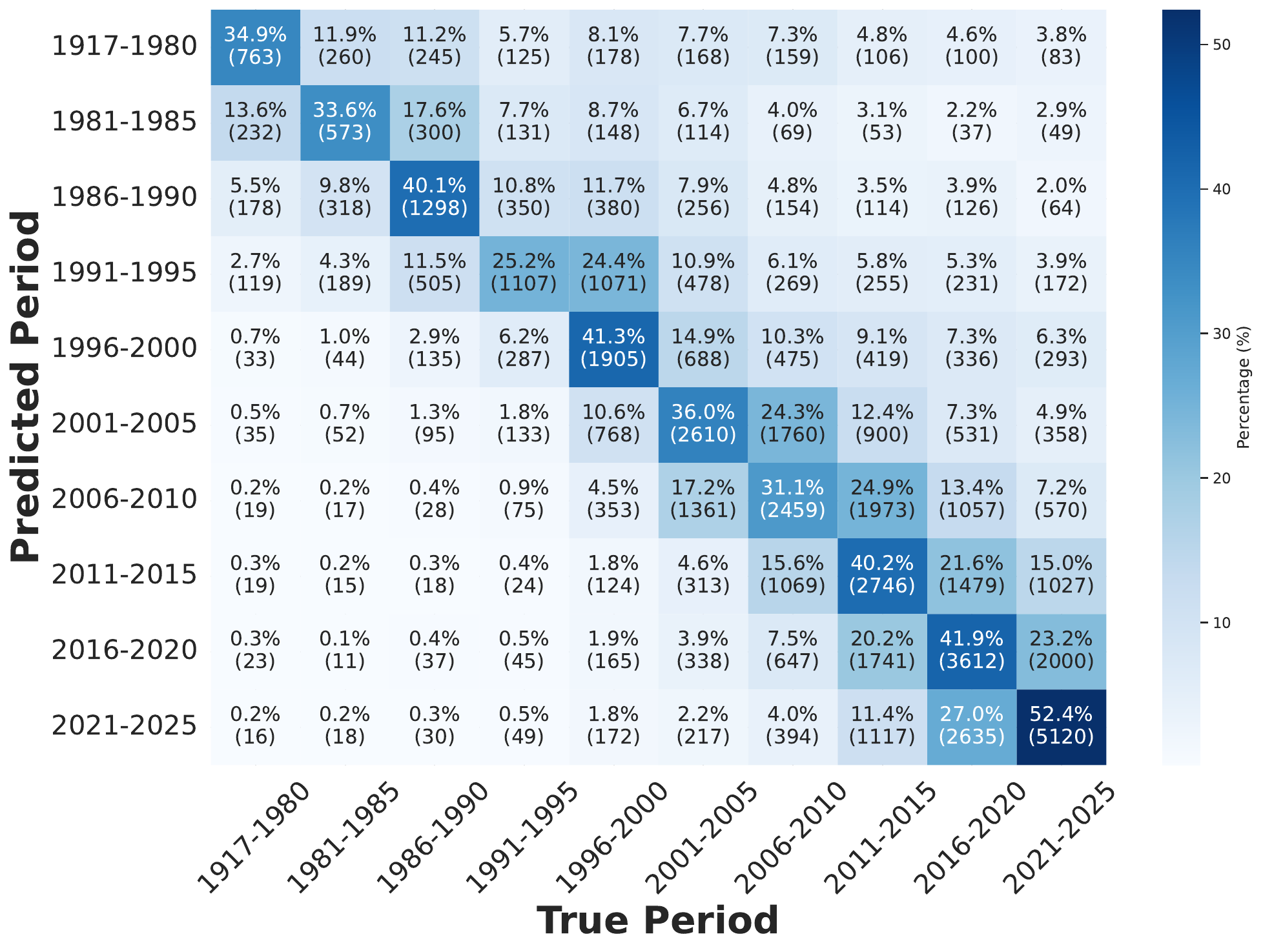}
        \caption{Confusion matrix of period prediction for the original avatars, showing that visual features can distinguish temporal periods, with most errors between adjacent periods.}
        \label{fig:period-predict-confusion-mat}
\end{figure}

\section{Confusion Matrix for Personality Archetype Prediction}

\begin{figure}[H]
  \centering
  \includegraphics[width=0.48\textwidth]{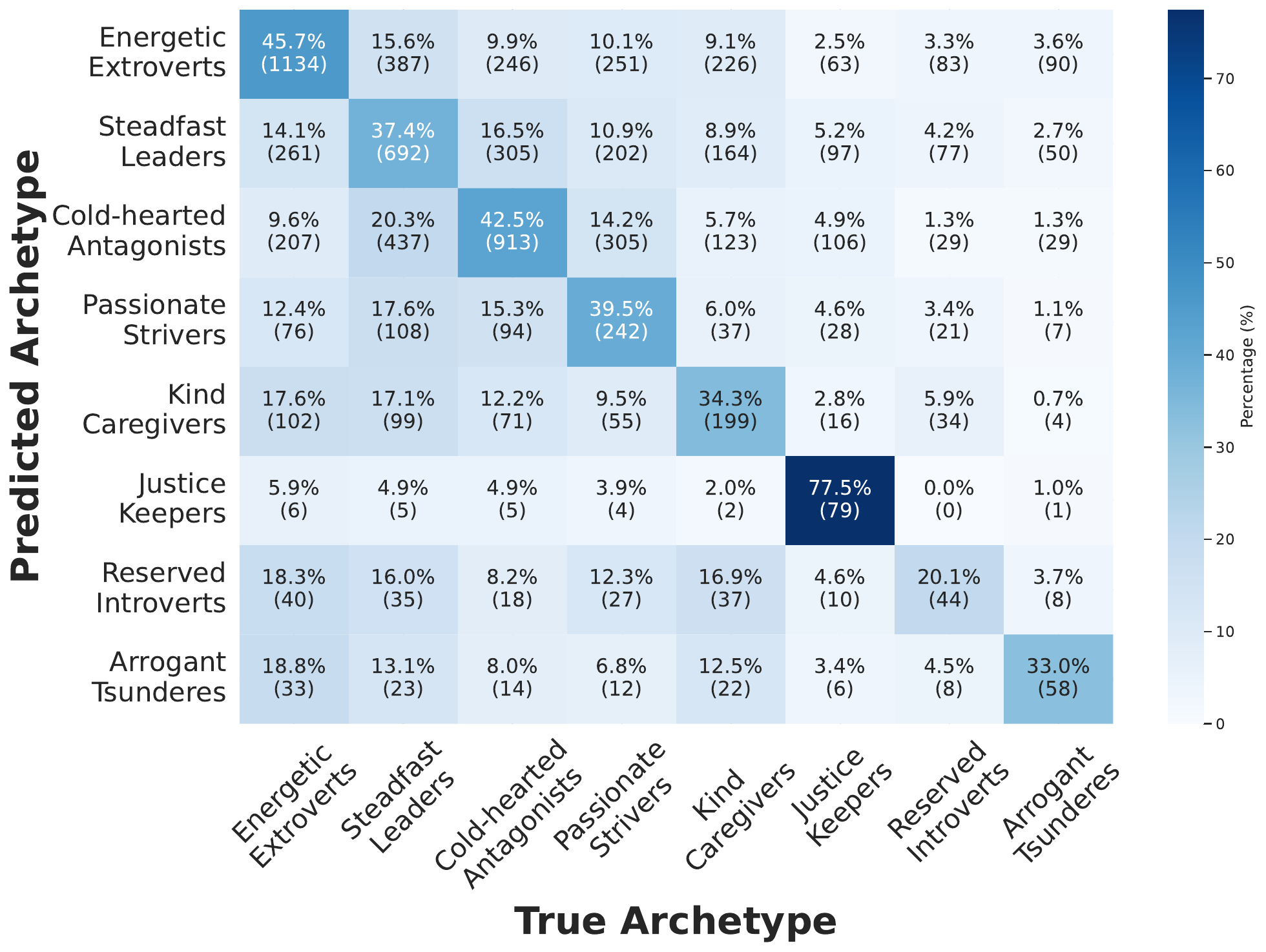}
  \caption{Confusion matrix for personality archetype prediction from the first 100 principal components of DINOv2 avatar embeddings. Justice Keepers, Energetic Extroverts, and Cold-hearted Antagonists are most predictable.}
  \label{fig:pc-comm-pred}
\end{figure}

\section{Full Results for User Preference Modeling}  \label{appdx:pred-pref-results}

\begin{table}[H]
\centering
\begin{tabular}{lllcc}
\toprule
\textbf{Features}  & \textbf{PCs}  & \textbf{Model}  & \textbf{MAE}  & \textbf{MPD} \\
\midrule
Baseline (mean)  & --  & --  & 1.190  & 2.166 \\
\midrule
Both PCs  & 20  & Poisson  & 0.984  & 1.640 \\
Both PCs  & 20  & HGBT  & 0.881  & 1.354 \\
Both PCs  & 50  & Poisson  & 0.963  & 1.583 \\
Both PCs  & 50  & HGBT  & 0.880  & 1.345 \\
Both PCs  & 100  & Poisson  & 0.941  & 1.522 \\
Both PCs  & 100  & HGBT  & 0.879  & 1.340 \\
\midrule
Both PCs + Meta  & 20  & Poisson  & 0.897  & 1.416 \\
Both PCs + Meta  & 20  & HGBT  & 0.669  & 0.946 \\
Both PCs + Meta  & 50  & Poisson  & 0.886  & 1.385 \\
Both PCs + Meta  & 50  & HGBT  & \textbf{0.668}  & \textbf{0.945} \\
Both PCs + Meta  & 100  & Poisson  & 0.868  & 1.343 \\
Both PCs + Meta  & 100  & HGBT  & 0.669  & 0.946 \\
\midrule
Avatar  & 20  & Poisson  & 1.022  & 1.725 \\
Avatar  & 20  & HGBT  & 1.015  & 1.693 \\
Avatar  & 50  & Poisson  & 1.001  & 1.666 \\
Avatar  & 50  & HGBT  & 1.008  & 1.673 \\
Avatar  & 100  & Poisson  & 0.978  & 1.601 \\
Avatar  & 100  & HGBT  & 1.007  & 1.663 \\
\midrule
Avatar + Meta  & 20  & Poisson  & 0.921  & 1.461 \\
Avatar + Meta  & 20  & HGBT  & 0.724  & 1.071 \\
Avatar + Meta  & 50  & Poisson  & 0.911  & 1.432 \\
Avatar + Meta  & 50  & HGBT  & 0.723  & 1.069 \\
Avatar + Meta  & 100  & Poisson  & 0.893  & 1.387 \\
Avatar + Meta  & 100  & HGBT  & 0.724  & 1.069 \\
\midrule
Personality  & 20  & Poisson  & 1.119  & 1.997 \\
Personality  & 20  & HGBT  & 0.972  & 1.590 \\
Personality  & 50  & Poisson  & 1.110  & 1.979 \\
Personality  & 50  & HGBT  & 0.970  & 1.585 \\
Personality  & 100  & Poisson  & 1.102  & 1.964 \\
Personality  & 100  & HGBT  & 0.969  & 1.583 \\
\midrule
Personality + Meta  & 20  & Poisson  & 0.958  & 1.575 \\
Personality + Meta  & 20  & HGBT  & 0.699  & 1.022 \\
Personality + Meta  & 50  & Poisson  & 0.953  & 1.564 \\
Personality + Meta  & 50  & HGBT  & 0.699  & 1.022 \\
Personality + Meta  & 100  & Poisson  & 0.950  & 1.558 \\
Personality + Meta  & 100  & HGBT  & 0.699  & 1.022 \\
\midrule
Meta only  & --  & Poisson  & 0.994  & 1.651 \\
Meta only  & --  & HGBT  & 0.779  & 1.204 \\
\bottomrule
\end{tabular}
\caption{Full results of predicting log-transformed favorite counts using avatar PCs, personality PCs, and metadata (anime favorites, gender, role, year).}
\label{tab:fav_pred_full_results}
\end{table}

\end{document}